\documentclass[%
 reprint,
showpacs,
 amsmath,amssymb,
 aps,
 pra,
floatfix,
superscriptaddress
]{revtex4-1}
\usepackage[utf8]{inputenc}
\usepackage[T1]{fontenc}
\usepackage[]{graphicx}
\usepackage{color}
\usepackage[dvipsnames]{xcolor}
\usepackage[english]{babel}
\usepackage[normalem]{ulem}
\usepackage[bookmarks,bookmarksopen,bookmarksdepth=2]{hyperref}
\hypersetup{
    colorlinks=true,
    citecolor=blue,
    linkcolor=blue,
    filecolor=magenta,
    urlcolor=cyan,
}

\usepackage{microtype}
\usepackage{braket}
\usepackage{esvect}
\usepackage[export]{adjustbox}
\usepackage{dsfont}
\usepackage[T1]{fontenc}
\usepackage[utf8]{inputenc}
\usepackage{url}
\usepackage{orcidlink}

\usepackage{amsmath,amsfonts,amssymb,amsthm}
\usepackage{mathtools}
\usepackage{bbm}
\usepackage{mathrsfs,physics,xfrac,cancel,tensor,relsize,tikz,lipsum,quantikz}
\usepackage[draft,inline,nomargin]{fixme}

\usepackage{xcolor}
\definecolor{darkgreen}{rgb}{0.0, 0.5, 0.0}

\newcommand{\medketbrasingle}[1]
{
\big|#1\big\rangle\big\langle#1\big|
}

\newcommand{\medket}[1]
{
\big|#1\big\rangle
}

\newcommand{\medmel}[3]
{
\big\langle#1\big|#2\big|#3\big\rangle
}

\begin{document}

\title{Fragility of Optimal Measurements due to Noise in Probe States for Quantum Sensing}
\author{Andrew Kolmer Forbes\orcidlink{0009-0003-8730-8007}}
\email{aforbes@unm.edu}
\affiliation{Center for Quantum Information and Control, Department of Physics and Astronomy, University of New Mexico, Albuquerque, New Mexico 87131, USA}

\author{Marco A. Rodríguez-García\orcidlink{0000-0003-1504-0526} }
\affiliation{Center for Quantum Information and Control, Department of Physics and Astronomy, University of New Mexico, Albuquerque, New Mexico 87131, USA}
\affiliation{Institute for Quantum Science and Technology, University of Calgary, Calgary, AB T2N 1N4, Canada}

\author{Ivan H. Deutsch}
\affiliation{Center for Quantum Information and Control, Department of Physics and Astronomy, University of New Mexico, Albuquerque, New Mexico 87131, USA}

\begin{abstract}
For a given quantum state used in sensing, the quantum Cramér-Rao bound (QCRB) sets a fundamental limit on the precision achievable by an unbiased estimator of an unknown parameter, determined by the inverse of the quantum Fisher information (QFI). The QFI serves as an upper bound on the classical Fisher information (CFI), representing the maximum extractable information about the unknown parameter from measurements on a physical system. Thus, a central goal in quantum parameter estimation is to find a measurement, described by a POVM, that saturates the QFI (achieves maximum CFI), and thereby achieves the QCRB. In the idealization that one uses pure states and unitary encodings for sensing, discontinuities can appear in the CFI but not the QFI. In this article, we demonstrate that these discontinuities are important features, quantifying how much Fisher information is lost in the presence of noise.  We refer to this as the Fisher information ``fragility''. We present a simple framework for understanding how discontinuities increase fragility through Jensen’s inequality, and demonstrate how one can use this framework to design more robust POVMs for quantum advantage in metrology.  
\end{abstract}

\maketitle

\section{Introduction}

Quantum metrology, which allows for precision measurement beyond the standard quantum limit, has made significant advances in the past decades ~\cite{Braunstein1994,Giovannetti2011,Colombo2022,Toth2014,Schmidt2005,Degen2017}, with practical applications ranging from gravitational wave detection \cite{Tse2019} to
high-precision magnetometry \cite{Wang2025, Hans2024,Fiderer2018}, and dark matter searches~\cite{backes2021,Brady2022}. These advances rely on the use of nonclassical probe states,
which can improve the precision of parameter estimation beyond what is
achievable with classical resources \cite{Pezze2018,Forbes2025,Kwon2019}. The ultimate precision limit for parameter estimation in linear quantum
sensing, known as the Heisenberg limit, can be attained using entangled states
and reflects the optimal scaling with resources allowed by quantum mechanics.
This represents a quadratic improvement over the standard quantum limit, which
bounds the precision achievable with classical resources
\cite{Buzek1999,Giovannetti2006}.

In single-parameter estimation, the fundamental task is to estimate an
unknown parameter $\theta$ that is encoded in a quantum state
$\hat\rho_\theta$. A key challenge in quantum metrology is determining the
optimal measurement strategy to estimate $\theta$. A given measurement defined by a POVM determines a probability distribution of measurement outcomes, from which one can often construct an unbiased (or asymptotically unbiased) estimator $\hat{\theta}$.  Given $n$ samples from the probability distribution, the Cram\'er-Rao bound gives a lower bound on the estimator variance $\Delta \hat{\theta}^2\geq 1/(n \mathcal{F}_C(\theta))$, where $\mathcal{F}_C(\theta)$ is the classical Fisher information (CFI).  The quantum Fisher Information (QFI) is the maximum Fisher information one can obtain, maximized over all possible POVMs.  The optimal POVM is not necessarily unique.

Of particular importance is the effect of noise and decoherence in quantum metrology, which will tend to cancel out any quantum advantage.  A significant body of research in recent years has focused on identifying the maximum achievable sensitivity under realistic conditions, particularly in the presence of noise in the probe state or parameter encoding~\cite{Kurdzialek2023,Sekatski2017,Dobrzaski2017,Zhou2018}, as well as
imperfections in the measurement process \cite{Zhou2023,Len2022}. Recently, work from the group of Demkowicz-Dobrzański has derived a quantity aimed at capturing the susceptibility of the Fisher information to measurement noise \cite{Kurdzialek2023_FiMeNoS}. Aptly named the Fisher Information Measurement Noise Susceptibility (FI-MeNoS), this quantity is equal to the maximum rate of loss of Fisher information when the measurement corresponds to a convex combination of an intended POVM $\mathcal M$ and an unintended (noise) POVM $\mathcal N$. By treating the loss of Fisher information as a cost function, and maximizing this function over $\mathcal N$, they derive an explicit expression for the FI-MeNoS $\chi[\mathcal M]$. 

In this work we study which optimal POVMs are fragile in the face of noise, and which are robust.  We take a slightly different approach than the study of FI-MeNoS. Instead of considering a completely noiseless parameter encoding and noisy measurement, we consider the problem of noisy state preparation. We show that measurements which are considered susceptible in the FI-MeNoS formalism are also considered fragile in our formalism with respect to noisy input states. While the FI-MeNoS quantifies the rate of loss of Fisher information for ``worst possible'' measurement noise, our formalism focuses on understanding the approximate amount of Fisher information lost in the presence of very general noise applied to the initial probe. We also focus on developing simple heuristic tools to understand Fisher information fragility. We further show that by considering noise in the probe state, one can interpret these fragile measurements by studying what we refer to as \textit{points of discontinuity}, which are measurements for which the probability of an outcome goes to zero, but the contribution to the Fisher information remains finite. The connection between points of discontinuity and Fisher information fragility was pointed out in \cite{Kurdzialek2023_FiMeNoS}, and in this paper, we will to make this connection rigorous and intuitive. 

The remainder of this paper is organized as follows. In Sec.~\ref{sec:formalism} we provide the theoretical background necessary for
understanding our work. In Sec.~\ref{sec:examples} we
provide a motivating example that serves as a guiding understanding of our
analysis. In Sec.~\ref{sec:noise} we provide interpretations of why
discontinuities lead to a loss in Fisher information for noisy probe states. We demonstrate that fragility of Fisher information is a feature of many noise models, and derive pathological noise models for which discontinuities do not lead to fragility. Finally, in Sec.~\ref{sec:discussion} we discuss the large
system limit, relating large spin ensembles to bosonic modes, and analyze the bias of the maximum likelihood estimator
for a particular system and protocol.






\section{Fisher Information}
 \label{sec:formalism}

Let $\Lambda$ denote the classical outcome of a measurement. Its possible values
form a set $\mathcal S$ (discrete or continuous), and we model the outcome
statistics by a parametric family of probability distributions
$\{P_\theta:\theta\in\Theta\subseteq\mathbb R\}$ on $\mathcal S$. We assume that
the model admits a common reference measure $\mu$ (counting measure for discrete
outcomes, Lebesgue measure for continuous ones), so that each $P_\theta$ has a
density (or probability mass function) $p_\theta$ with respect to $\mu$, meaning
that $P_\theta(A)=\int_A p_\theta(\lambda)\,\mu(d\lambda), \,
A\subseteq\mathcal S$. For some unknown but fixed $\theta\in\Theta$, the
measurement outcome $\Lambda$ is distributed according to $P_\theta$, and the task
is to infer $\theta$ from observations of $\Lambda$.

The Fisher information $\mathcal F_C[p_\theta(\Lambda);\theta]$ about $\theta$
is defined as the variance, under $P_\theta$, of the score function
$s_\theta(\lambda)=\partial_\theta\log p_\theta(\lambda)$, and it quantifies how
sensitively the statistics $p_\theta(\lambda)$ depend on $\theta$. In
particular, when the outcome space is discrete, one has
\begin{align}
  \mathcal F_C[p_\theta(\Lambda);\theta]
  &= \sum_{\lambda\in\mathcal S_\theta}
     p_\theta(\lambda)\left(\partial_\theta \log p_\theta(\lambda)\right)^{2} \notag \\
  &= \sum_{\lambda\in\mathcal S_\theta}
     \frac{\big(\partial_\theta p_\theta(\lambda)\big)^{2}}{p_\theta(\lambda)}.
  \label{eq:discrete_cfi}
\end{align}
Here
\[
\mathcal S_\theta=\{\lambda\in\mathcal S \mid p_\theta(\lambda)>0\}
\]
denotes the support of $p_\theta$, ensuring the expressions are well-defined.
In the quantum setting considered below, we will focus on discrete outcome
distributions.

The Fisher information sets a lower bound on the variance of any unbiased
estimator $\hat\theta: \mathcal S^n \to \Theta$ based on $n$ independent and identically distributed
realizations of the random variable $\Lambda$, via the Cramér--Rao bound
\begin{equation}
\Delta\hat\theta^2\geq\frac1{n\mathcal F_C[p_\theta(\Lambda);\theta]}.
\end{equation}
Thus, the Fisher information quantifies the 
achievable precision of classical estimation strategies in such sensing protocols.

When sensing using a quantum state, it is often the case that $\theta$ is encoded by some unitary operator $\hat U_\theta$ generated by a Hamiltonian $\hat H$. The state $\hat \rho_\theta$ is then given by
\begin{equation}
    \hat \rho_\theta=e^{-i\theta \hat H}\hat \rho e^{i\theta \hat H},
\end{equation}
where $\hat \rho$ is the initial state. The probability distribution as a function of $\theta$ for some POVM, $\{\hat E_\lambda\}$, is then
\begin{equation}
    p_\theta(\lambda)=\Tr(\hat E_\lambda e^{-i\theta \hat H}\hat \rho e^{i\theta \hat H}).
\end{equation}
Written in terms of the quantum state $\hat \rho_\theta$ for the case $\theta=0$, the Fisher information, now expressed as a function of the POVM $\{\hat E_\lambda\}$ is 

\begin{equation}
    \mathcal F_C[\hat \rho;\hat E_\lambda]=-\sum_{\lambda\in \mathcal S_{\theta=0}}\frac{\Tr(\hat E_\lambda[\hat H,\hat \rho])^2}{\Tr(\hat E_\lambda\hat \rho)}.\label{eq:cfi_for_state}
\end{equation}
In doing so we obtain an expression for the Fisher information of the initial state $\hat \rho$. 
 

\subsection{Discontinuities in the Fisher information}
\label{sec:discontinuities}




Many desirable properties of the Fisher information as a function of $\theta$,
such as smoothness, rely on mild regularity conditions on the statistical model
$\{p_\theta : \theta \in \Theta\}$ introduced in Sec.~\ref{sec:formalism}. We
now fix a point $\theta' \in \Theta$ and work under the following standing
hypotheses, stated informally. (i) The support
$\mathcal S_\theta = \{\lambda \in \mathcal S : p_\theta(\lambda)>0\}$ is
locally constant in $\theta$ (or, more generally, any $\theta$-dependence of
the support is such that no boundary terms arise when differentiating the sums
in Eq.~(\ref{eq:discrete_cfi})). (ii) The map
$\theta \mapsto p_\theta(\lambda)$ is differentiable for almost every
$\lambda \in \mathcal S_{\theta'}$, with $\partial_\theta p_\theta$ continuous
in $\theta$. (iii) Differentiation may be interchanged with
integration/summation thanks to an $L^1$ envelope (in particular, this implies
$\mathbb E_\theta[s_\theta(\Lambda)]=0$). Finally, the score is
square-integrable at $\theta'$, i.e.\
$\mathbb E_{\theta'}[s_{\theta'}(\Lambda)^2]<\infty$. Under these conditions,
the Fisher information exists at $\theta'$ and satisfies $\mathcal F_C[p_\theta(\Lambda);\theta'] =
  \mathbb E_{\theta'}[s_{\theta'}(\Lambda)^2]$ (see, e.g., \cite{Lehman}).

The phenomenon of interest arises when the first of these conditions fails. If
the support $\mathcal S_\theta$ changes with $\theta$, then evaluating $\mathcal
F_C$ at $\theta'$ necessarily excludes outcomes $\lambda$ with
$p_{\theta'}(\lambda)=0$, whereas the limit $\theta\to\theta'$ can still
capture singular contributions $(\partial_\theta
p_\theta(\lambda))^2/p_\theta(\lambda)$ appearing in
Eq.~(\ref{eq:discrete_cfi}). Thus, the pointwise value at $\theta'$ need not agree with the limiting value,
and a discontinuity may ensue. In the favorable case that there exists $\lambda$ with
$p_{\theta'}(\lambda)=0$ and $\partial_\theta
p_\theta(\lambda)\big|_{\theta=\theta'}=0$, the singular ratio admits a
finite limit. Applying L’H\^opital’s rule, one finds that the size of the jump
  discontinuity of $\mathcal F_C$ at $\theta'$ is
\begin{align}
\Delta\mathcal F_C[p_\theta(\Lambda),\theta']
&= \lim_{\theta\to\theta'}\mathcal F_C[p_\theta(\Lambda);\theta]
       - \mathcal F_C[p_\theta(\Lambda);\theta'] \nonumber\\
&= 2\,\partial_\theta^2 p_\theta(\lambda)\Big|_{\theta=\theta'}.
\label{eq:disc_at_lambda}
\end{align}
Consequently, for a finite set $\mathcal
N_\theta=\{\lambda:\,p_{\theta'}(\lambda)=0\}$, the jump discontinuity is the sum of
the individual contributions, \( \Delta\mathcal F_C[p_\theta(\Lambda),\theta']
= 2\sum_{\lambda\in\mathcal N_{\theta'}}\partial_\theta^2
p_\theta(\lambda)\big|_{\theta=\theta'}. \)

To illustrate this, consider $\Lambda=\{0,1\}$ a Bernoulli random variable and, for
$\theta$ near $0$, set $p_\theta(1)=\theta^2$ and $p_\theta(0)=1-\theta^2$. At
$\theta'=0$ the outcome $\lambda=1$ has zero probability, and moreover
$\partial_\theta p_\theta(1)\big|_{\theta=0}=0$ while $\partial_\theta^2
p_\theta(1)\big|_{\theta=0}=2$. For $\theta\neq 0$ one computes
\[
\mathcal F_C[p_\theta(\Lambda);\theta]
=\frac{(2\theta)^2}{\theta^2}+\frac{(-2\theta)^2}{1-\theta^2}
=4+\frac{4\theta^2}{1-\theta^2}\xrightarrow[\theta\to 0]{}4,
\]
whereas at $\theta'$ the term for $\lambda=1$ is excluded and
\[
\mathcal F_C[p_\theta(\Lambda);\theta']
=\frac{\big(\partial_\theta p_\theta(0)\big|_{\theta=0}\big)^2}{p_0(0)}
=\frac{0^2}{1}=0.
\]
Hence $\Delta\mathcal F_C[p_\theta(\Lambda),\theta_0]=4$, in agreement with
(\ref{eq:disc_at_lambda}) since $2\,\partial_\theta^2
p_\theta(1)\big|_{\theta=0}=4$.


In contrast, if \(p_{\theta'}(\lambda)=0\) but \(\partial_\theta p_\theta(\lambda)\big|_{\theta=\theta'}\neq 0\). Taylor’s expansion around \(\theta'\) then
yields
\[
p_\theta(\lambda)
= \partial_\theta p_\theta(\lambda)\big|_{\theta'}\,(\theta-\theta')
  + o(|\theta-\theta'|),
\]
so the contribution of this $\lambda$ to the Fisher information satisfies
\[
\frac{\big(\partial_\theta p_\theta(\lambda)\big)^2}{p_\theta(\lambda)}
\sim \frac{\partial_\theta p_\theta(\lambda)\big|_{\theta'}}{\,|\theta-\theta'|\,}
\qquad(\theta\to\theta').
\]
Hence, the limit blows up, whereas at $\theta'$ the $\lambda$ contribution vanishes. Therefore, $\mathcal F_C$ thus has an infinite discontinuity at $\theta'$.  However, in many
estimation problems with quantum systems, the first derivative at $\theta'$ vanishes by analyticity, so the zero is at least quadratic and the contribution is finite, as captured by Eq.~\eqref{eq:disc_at_lambda}.

For a quantum state $\hat \rho_\theta$, where $\theta$ is encoded by a unitary map 
generated by $\hat H$, one may write the size of the discontinuity at $\theta=\theta'$
as 
\begin{align} \Delta\mathcal F_C[\hat \rho_{\theta'};\hat
  E_\lambda]=-2\sum_{\lambda\in \mathcal N_{\theta'}}\Tr(\hat E_\lambda [\hat H,[\hat
  H,\hat \rho_{\theta'}]]).\label{eq:disc_mixed} \end{align} 
For pure states, it follows from the lemma proven in App.~\ref{app:trace_proof} that
\begin{align} \Delta\mathcal F_C[\hat \rho_{\theta'};\hat
  E_\lambda]=4\sum_{\lambda\in \mathcal N_{\theta'}}\Tr(\hat E_\lambda \hat H\hat
  \rho_{\theta'} \hat H).\label{eq:disc_main} 
  \end{align}
When the POVM is a set of rank-1
projectors $\{\hat E_\lambda=\ketbra\lambda \}$, one may further simplify this to
\begin{equation} 
\label{eq:DeltF}
\Delta\mathcal F_C[\hat \rho_{\theta'};\hat
  E_\lambda]=4\sum_{\lambda\in \mathcal N_{\theta'}}|\mel{\lambda}{\hat
    H}{\psi_{\theta'}}|^2   
\end{equation} 
where  $\hat\rho_{\theta'}=\ketbra{\psi_{\theta'}}$. For the remainder of this article, we
will focus on discontinuities in the context of pure states and unitary
encodings of $\theta$, and thus we will refer back to Eq.~(\ref{eq:DeltF})
as the size of the discontinuity in the Fisher information.


In this work we use the above discussion to investigate the fragility of a
  measurement scheme, by which we mean the appearance of large changes in the
  classical Fisher information under arbitrarily small perturbations of the state. Concretely, we consider a unitarily equivalent family of
  POVMs $\{\hat E_{\lambda}^{(\beta)}\}$ obtained by rotating a
  fiducial measurement $\{\hat E_\lambda\}$,
\[
  \hat E_{\lambda}^{(\beta)}=\hat U_\beta \hat E_\lambda \hat U_\beta^\dagger,
\]
where $\beta\in\mathbb R$ parametrizes a controllable unitary acting on the
measurement. The corresponding outcome probabilities are
$p_{\theta,\beta}(\lambda)=\Tr(\hat\rho_\theta \hat E_{\lambda}^{(\beta)})$. By
cyclicity of the trace,
\[
p_{\theta,\beta}(\lambda)
=\Tr(\hat U_\beta^\dagger \hat\rho_\theta \hat U_\beta\,\hat E_\lambda),
\]
so varying $\beta$ at fixed $\theta$ is equivalent to keeping the fiducial POVM
fixed and applying the inverse unitary to the state prior to measurement.

We therefore treat $\beta$ as a control parameter selecting the measurement,
  and study the Fisher information about $\theta$ at fixed $\theta=\theta'$ as a
  function of $\beta$. For pure probe states, the measurement is fragile near those values of
  $\beta$ for which one or more ideal outcome probabilities vanish at $\theta'$,
  that is, when $p_{\theta',\beta}(\lambda)=0$ for some $\lambda$. At such
  points the support changes and $\mathcal F_C$ can be discontinuous as a
  function of $\beta$, with the jump determined by the preceding analysis,
  equivalently by Eq.~(\ref{eq:disc_at_lambda}) applied to
  $p_{\theta,\beta}(\lambda)$, and by Eq.~(\ref{eq:DeltF}) in the pure-state
  unitary-encoding setting.

\subsection{Quantum Fisher Information}

The quantum extension of Fisher information is the so-called \emph{quantum Fisher information} (QFI) $\mathcal F_Q[\hat \rho_\theta ;\theta]$. It is the maximum achievable Fisher information attainable by measuring a state $\hat \rho_\theta$ using an optimal POVM. This leads to the quantum Cram\'er-Rao bound (QCRB) which bounds the ultimate achievable variance of an unbiased estimator $\hat\theta$ for sensing using a quantum state $\hat \rho_\theta$
\begin{equation}
    \Delta\hat\theta^2\geq\frac1{n\mathcal F_Q[\hat \rho_\theta ;\theta]}.\label{eq:QCRB}
\end{equation}
In analogy to how the Fisher information is the variance of the score function, the QFI is the variance of the quantum analog of the score function, known as the \textit{symmetric logarithmic derivative}, $\hat L$, which is defined implicitly by
\begin{equation}
    \frac{\partial}{\partial\theta}\hat \rho_\theta=\frac12\big(\hat \rho_\theta \hat L_\theta+\hat L_\theta\hat \rho_\theta\big).
\end{equation}
The QFI is then 
\begin{equation}
    \mathcal F_Q[\hat \rho_\theta;\theta]=\Tr(\hat \rho_\theta \hat L_\theta^2),
\end{equation}
since the definition of $\hat L_\theta$ implies that $\Tr(\hat \rho_\theta \hat L_\theta)=0$. 

An ``optimal measurement'' in quantum sensing often refers to a POVM for which the Fisher information (which from here on we refer to as the \textit{classical} Fisher information) is equal to the quantum Fisher information. Finding such a measurement implies that in the asymptotic limit of taking many samples from $\hat \rho_\theta$, one can saturate the QCRB, provided one is able to find an efficient, unbiased estimator. For a family of optimal measurements, our goal is to determine which are robust in the face of noise, and which are fragile.  We will show in the following sections that fragility occurs near discontinuities in the classical Fisher information.

\begin{figure*}[t]
    \centering
    \includegraphics[width=0.94\linewidth]{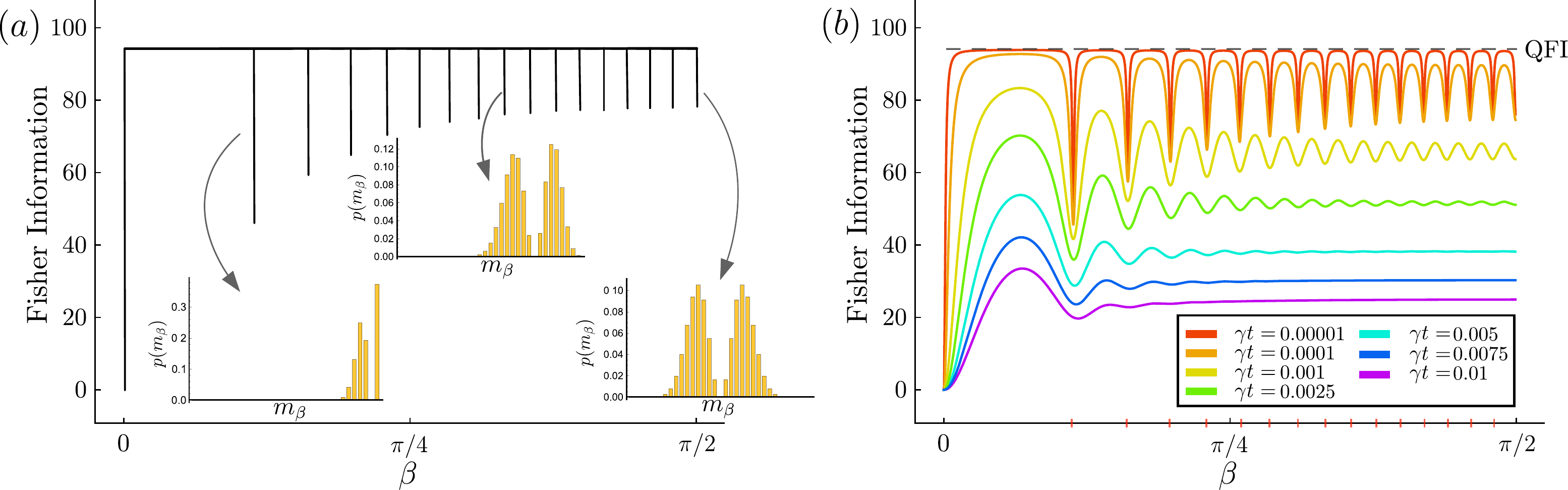}
    \caption{Classical Fisher information (CFI) for (a) a pure initial first excited Dicke state, with $J=16$, Eq.~(\ref{eq:initial_state}) as a function of $\beta$, the angle of the measurement basis given by the eigenstates of $\exp\small(-i\beta\hat J_y\small)\hat J_z \exp\small(i\beta\hat J_y\small)$. The CFI in (a) is independent of $\beta$ except on a set of measure zero corresponding POVMs where the probability of some outcomes is zero (see insert).  At these values of $\beta$ the CFI discontinuously jumps to a lower value. (b) The CFI as a function of $\beta$ after the probe state is put through a collective depolarizing channel, Eq.~(\ref{eq:collective_lindbladian}), for different amounts of noise $\gamma t$.  The dashed line represents the quantum Fisher information (QFI) which is the maximum achievable Fisher information over all possible measurements.  For the noisy probe state, the CFI is continuous, dipping near the points of discontinuity in (a).  Measurement bases near these values of $\beta$, while optimal for pure states, are ``fragile" in presence of noise.}
    \label{fig:w_state}
\end{figure*}

\section{Example of Fragility of the Fisher Information in Spin Systems}
\label{sec:examples}

The effects of noise and the fragility of measurements are seen clearly in a basic example  -- sensing a rotation on the Bloch sphere with an ensemble of $N$ qubits.  We consider the scenario in which all $N$ qubits sense the same field, corresponding to a rotation of the collective spin $\hat{\mathbf{J}} = \sum_i \hat{\vec{\sigma}}^{(i)}/2$, with total angular momentum $J=N/2$.  We sense the rotation around the $y$-axis $\hat{U}_\theta =e^{-i\theta \hat{J}_y}$.  For an initial unentangled spin coherent state $\ket{\uparrow_z}^{\otimes{N}} = |J, M= J\rangle$ the uncertainty in estimating $\theta$ set by the QCRB scales as $1/\sqrt{N}$, the standard quantum limit (SQL). Dicke states,  $\ket{J,M}$ with $|M|<J$, are entangled states that have been proposed as sensors for rotations to achieve sensitivity advantage over the SQL \cite{Lee2025,Zhang2014,Sanders1995,Holland1993,Saleem2024,Zou2018}. In the absence of noise,  the first lower Dicke state $|J, M= J-1 \rangle$, equivalent to a a W-state of the $N$ qubits, achieves a constant factor improvement in the Fisher information compared to the spin coherent state. We will use this state in the following example.

To sense the rotation, one ideally prepares the state
\begin{align}
    \hat \rho_\theta=e^{-i\theta \hat J_y}\ketbra{J,J-1}e^{i\theta \hat J_y}.\label{eq:initial_state}
\end{align}
The QFI for this state is $6J-2$ \cite{Pezze2018}, approximately a three-times improvement over a spin coherent state.  Figure 1a shows the CFI for the family of POVMs corresponding to projective measurements in the basis $\ket{J,M_\beta}\equiv e^{-i\beta J_y}\ket{J,M}$, as function of $\beta$, for $N=32$, $J=16$.  We see that for all $\beta$ the CFI saturates the QFI save for a number of discrete points which make up a set of measure zero.  These correspond to points at which the the probability of some measurement outcome is zero, $\bra{J,M_\beta}\hat\rho_{\theta=0}\ket{J,M_\beta}=|\braket{J,M_\beta}{J,J-1}|^2=0$, and thus we expect discontinuities in the Fisher information.  As shown in App.~\ref{app:dicke_state_analysis} these points occur at $\beta=\arccos(M/J)$. The corresponding size of the discontinuity, Eq.~(\ref{eq:DeltF}) is (see App. \ref{app:dicke_state_analysis}),
\begin{align}
\label{eq:size_disc}
      \Delta \mathcal F_C=&4|\mel{J,M_\beta}{\hat J_y}{J,J-1}|^2 \\
    =&8J\left(\frac{J+M}{2J}\right)^{J+M}\left(\frac{J-M}{2J}\right)^{J-M}\binom{2J}{J-M}. \nonumber \
\end{align}
While in the ideal noiseless case, these discontinuities are negligible in the choice of the optimal POVM (as they occur on a set of measure zero within the domain of $\beta$), they have a profound impact in the presence of noise and decoherence. 

Consider the effect of collective Markovian noise, corresponding to the Lindbladian
\begin{equation}
\mathcal{L}\hat{\rho}=\mathcal{D}[\hat{J}_x]\hat{\rho}+\mathcal{D}[\hat{J}_y]\hat{\rho}+\mathcal{D}[\hat{J}_z]\hat{\rho},\label{eq:collective_lindbladian}
\end{equation}
where
\begin{equation}
\mathcal{D}[\hat{J}_i]\hat{\rho} =-\frac{\gamma}{2}[J_i,[J_i,\rho]].
\end{equation}
This corresponds to a ``collective depolarizing noise'', and should be visualized as the state spreading over the collective Bloch sphere at rate $\gamma$.   The superoperator map describing this channel is denoted $\hat\rho (t) =\exp\{\mathcal{L}t\}[\hat\rho]$.  
The probe state after application of this noise channel for time $t$ is
\begin{equation}
    \hat \rho_\theta(t)=e^{-i\theta \hat J_y}\left(e^{t\mathcal L}\left[\ketbra{J,J-1}\right] \right)e^{i\theta \hat J_y}.
\end{equation}
For the decohered state, we calculate the classical Fisher information for the family of measurement bases $\ket{J,M_\beta}$. As seen in Fig.~\ref{fig:w_state}b, the CFI dips substantially around the angles at which the discontinuities occurred.  Thus, these measurement bases are {\em fragile} and experience a decrease in CFI, by an amount comparable to magnitude  of the discontinuity in Eq. (\ref{eq:size_disc}). The values of $\beta$ at which the discontinuities occur are also points where the FI-MeNoS diverges, indicating that the rate of loss of Fisher information approaches infinity near these points.



This simple example helps highlight several important facts. The first is that there exists certain measurements and probe states for which a pure initial probe state may saturate the quantum Fisher information, and yet small amounts of noise can drastically reduce the classical Fisher information. Second, these ``fragile'' conditions occur when the support $\mathcal S_\theta$ changes. This begs the question, why do the effects of these discontinuities persist even when the the support is constant (which is often the case when the probe state is noisy)? One may attempt to use a pure probe state for a sensing task, for which probability on a particular outcome is small but nonzero. While the CFI might appear optimal in this case, we find that even small amounts of noise can cause the CFI to behave as if one were using the probe state at the point of discontinuity. In the next section we will explore several useful explanations of this behavior.



\section{Effects of Discontinuities in the Presence of Noise}\label{sec:noise}

In this section we will explore in detail several interpretations of the ``loss of Fisher information'' observed in the example in the previous section. We define the loss of Fisher information to be the difference between the Fisher information $\mathcal F_C[\hat\rho;\hat E_\lambda]$ for the initial pure probe state $\hat\rho$, and the Fisher information after some small amount of noise  has occurred, $\mathcal F_C[\hat\rho';\hat E_\lambda]$, where $\hat\rho'= \Phi[\hat\rho]$ for some CPTP map, $\Phi[\cdot]$. To simplify the discussion in the following sections we restrict our attention to projective measurements, and a family of POVMs $\{\hat E_\lambda^{(\beta)}\}$ which corresponds to a unitary rotation by $\beta$ of a fiducial measurement basis $\{\hat E_\lambda = \ketbra{\lambda}\}$.


\subsection{Discontinuities and Loss of Signal}
\label{sec:signal}

One can think about the loss of Fisher information from the perspective of the signal of a measurement,
\begin{equation}
    \mathbb E[\Lambda]=\sum_\lambda\lambda p_\theta(\lambda).
\end{equation}
We consider a decrease in the signal to noise ratio $\text{SNR}_\lambda$ associated with outcome $\lambda$, which we define as,
\begin{equation}
    \text{SNR}_\lambda\equiv\frac{(1-\epsilon)\lambda p(\lambda)}{\epsilon\lambda\sigma(\lambda)}=\frac{(1-\epsilon) p(\lambda)}{\epsilon\sigma(\lambda)}
\end{equation}
for some noise distribution, $\sigma(\lambda)$, which affects $p_\theta(\lambda)$ such that the new total probability distribution after noise is $P_{\theta}(\lambda)=(1-\epsilon)p_\theta(\lambda)+\epsilon\sigma(\lambda)$. Here the noise $\sigma(\lambda)$ is assumed to be independent of $\theta$.

The rate of change of the signal with respect to $\theta$ often informs the optimal ``operating point'' of a measurement, as one ideally measures in such a way that the signal is a strong function of $\theta$. However, it is important to note that a weak dependence on $\theta$ does not directly imply low Fisher information, as can be seen in the example using a pure qubit in App.~\ref{sec:qubit}. The case where the signal is zero is unique in this respect, as it directly implies zero Fisher information, which can be deduced through inspection of Eq.~(\ref{eq:discrete_cfi}). We note for completeness, however, that a lower bound on the Fisher information is obtained through the rate of change of the signal~\cite{Stein2014},
\begin{equation}
    \mathcal F_C[p_\theta(\Lambda);\theta]\geq\frac{\left(\frac{\mathrm d}{\mathrm d\theta}\mathbb E_\theta[\Lambda]\right)^2}{\Delta\Lambda^2}.\label{eq:signal_lower_bound}
\end{equation}
We further note that when the probability of outcome $\lambda$ goes to zero as a function of $\theta$, its derivative with respect to $\theta$ must also go to zero for smooth $p_\theta(\lambda)$, i.e.,
\begin{equation}
    \lim_{\theta\to\theta'}p_\theta(\lambda)=0\;\;\Rightarrow \;\;\lim_{\theta\to\theta'}\frac{\mathrm d}{\mathrm d\theta}p_\theta(\lambda)=0,
\end{equation}
which is a result of the derivative of $p_{\theta}(\lambda)$ existing for all $\theta$ and $p_\theta(\lambda)\geq0$. This fact was also present in our derivation of the discontinuity size in Sec.~\ref{sec:discontinuities}.

Given this, it is clear from Eq.~(\ref{eq:discrete_cfi}) that when the probability of $\lambda$ is zero, the contribution to the signal from $\lambda$ is also zero, and thus that term in the summation in Eq.~(\ref{eq:discrete_cfi}) no longer contributes the the total Fisher information. In the noiseless case, small but nonzero $p_\theta(\lambda)$ can still yield maximum Fisher information as Eq.~(\ref{eq:signal_lower_bound}) only sets a lower bound. 

To understand why noise affects otherwise optimal measurements so severely when near discontinuities, consider the contribution to the Fisher information from $\lambda$ when $p_{\theta'}(\lambda)\ll1$. We note that Eq. (\ref{eq:disc_at_lambda}) implies that
\begin{equation}
    \left(\frac{\partial}{\partial \theta}p_\theta(\lambda)\right)^2\bigg\vert_{\theta\approx\theta'}\approx p_\theta(\lambda)\Delta\mathcal F_C[p_\theta(\Lambda),\theta']\bigg\vert_{\theta\approx\theta'}.
\end{equation}
Using this, we consider the inclusion of some small $\theta$-independent noise $\sigma(\lambda)$ such that the new probability distribution is $(1-\epsilon)p_\theta(\lambda)+\epsilon \sigma(\lambda)$. The contribution to the Fisher information from $\lambda$ for $\theta\approx\theta'$ now becomes
\begin{align}
    \nonumber&\frac{(1-\epsilon)^2\left(\frac{\partial}{\partial \theta}p_\theta(\lambda)\right)^2}{(1-\epsilon)p_\theta(\lambda)+\epsilon\sigma(\lambda)}\\
    &\;\;\;\approx(1-\epsilon)\Delta\mathcal F_C[p_\theta(\Lambda);\theta]\left(\frac{(1-\epsilon)p_\theta(\lambda)}{\epsilon\sigma(\lambda)}\right)\\
    &\;\;\;=(1-\epsilon)\Delta\mathcal F_C[p_\theta(\Lambda);\theta]\text{SNR}_\lambda,\label{eq:snr_fi}
\end{align}
where in the last line we have identified the signal-to-noise ratio for just outcome $\lambda$, SNR$_\lambda$. Equation (\ref{eq:snr_fi}) can be understood as meaning that when $\epsilon\ll1$ and SNR$_\lambda$ is small ($\theta\approx\theta'$) the contribution to the total Fisher information from $\lambda$ is approximately 0. However, as $\theta$ moves farther from $\theta'$ the contribution to the CFI from $\lambda$ increases at a rate approximately proportional to the rate of increase of the signal to noise ratio on outcome $\lambda$.


Phrasing the loss in Fisher information in terms of signal-to-noise helps connect our results with practical implementations of the measurement. We have shown that near discontinuities, the contribution to the Fisher information can be limited by the signal to noise ratio on a particular measurement outcome. In systems with many possible measurement outcomes the SNR going to zero on a single measurement outcome may only change the Fisher information slightly, since the overall signal is not affected as much. This can be seen in the example in Sec.~\ref{sec:examples}. In that example when $\beta=0$ the loss in Fisher information is maximal, corresponding to a measurement where all but one outcome $\lambda$ have SNR tending to 0. Every other value of $\beta$ leading to a discontinuity only leads to a loss of signal on a single outcome $\lambda$, and thus the loss is less severe.

\subsection{Markovian Noise and Continuity of the Fisher Information}
\label{sec:markovian_noise}

We now analyze the Fisher information in a way similar to the previous section, but specialized to Markovian noise at short times. This allows us to understand the conditions under which noise removes some discontinuities of the Fisher information.

\begin{figure}
    \centering
    \includegraphics[width=.95\linewidth]{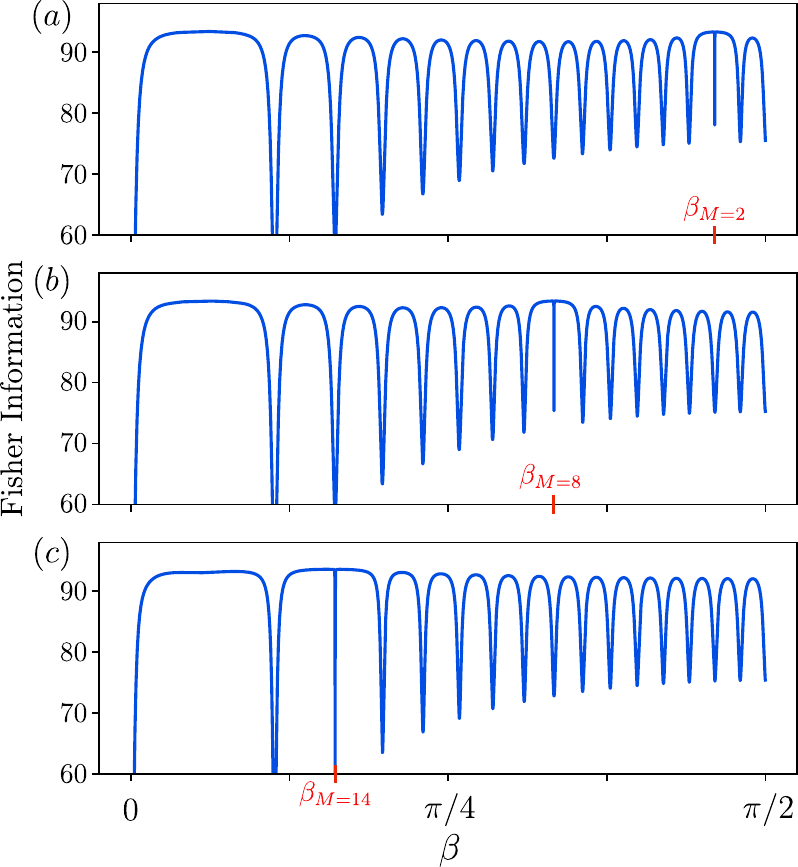}
    \caption{Fisher information at $\theta=0$ of a first excited Dicke state $\ket{J,J-1}$ for a $J=16$ spin system with respect to sensing rotations around the $\hat J_y$ axis parameterized by $\theta$. The Fisher information is calculated for measurements in the eigenbasis of $e^{-i\beta \hat J_y}\hat J_ze^{i\beta \hat J_y}$.  The noise is determined by the jump operator defined implicitly in Eqs.~(\ref{eq:pathological_condition}) and (\ref{eq:disc_special_case}) (constructed explicitly in App.~\ref{sec:explicit_construction}), applied at rate $\gamma$ for time $\gamma t=10^{-4}$, and is chosen such that one single discontinuity does not impact the Fisher information for values of $\beta$ surrounding it.  In (a), the jump operator $\hat L_{M=2}$ is constructed such that $M=2$ in Eq.~(\ref{eq:disc_special_case}), and thus the Fisher information surrounding the discontinuity at $\beta_{M=2}$ is unaffected by presence of the discontinuity. (b) and (c) show the same,  with $\hat L_{M=8}$ and $\hat L_{M=14}$ respectively.}
    \label{fig:removed_discontinuities}
\end{figure}

\begin{figure*}
    \centering
    \includegraphics[width=1\linewidth]{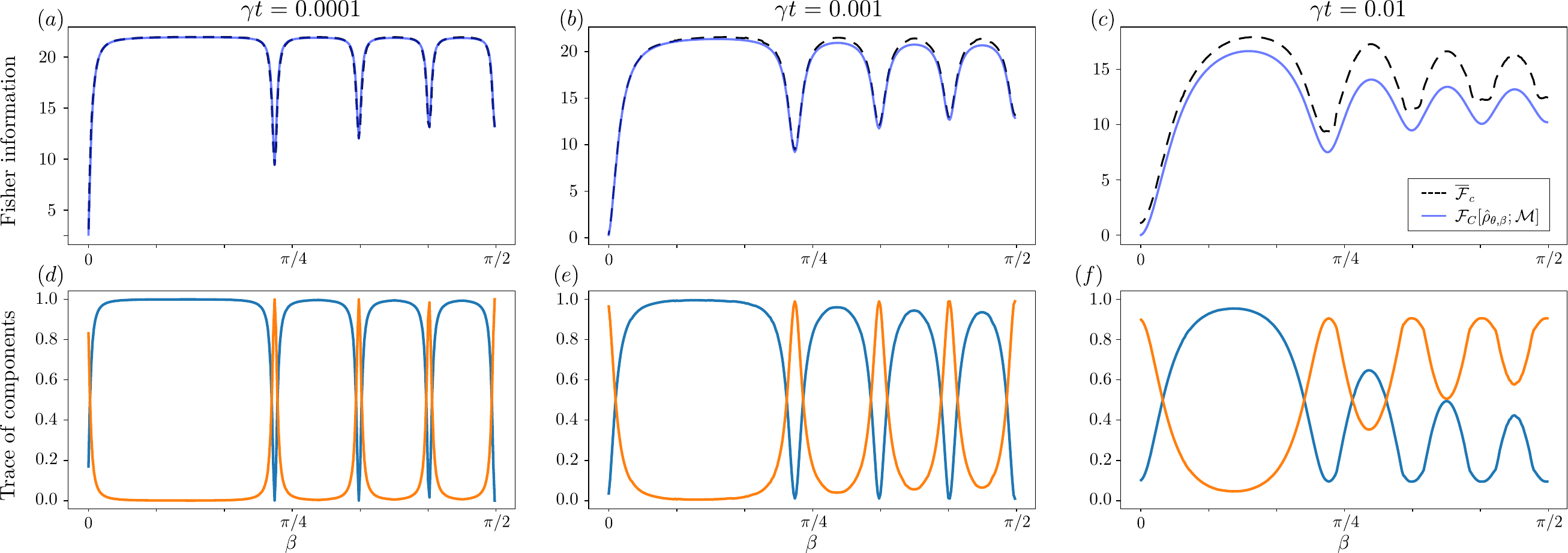}
    \caption{Top row (a-c): Classical Fisher information (blue) of the probe state corresponding a first excited Dicke state with $J=4$, after being subjected the noise channel, Eq.~(\ref{eq:collective_lindbladian}). The probe state is rotated about the $\hat J_y$ axis (after encoding of $\theta$) by angle $-\beta$ before being measured along the $\hat J_z$ axis. The dashed line is the Jensen's upper bound, Eq.~(\ref{eq:avg_fc}), determined from the pure state decomposition in Eq.~(\ref{eq:bad_decomp}). For weak noise, the bound is tight. Bottom row (d-f): traces of $\sum_iq_{i,\beta}\medketbrasingle{\phi_\theta^{(i)}}$ (orange) and $\hat \Omega_{\theta,\beta}$ (blue) of the decompositions in (a-c).}
    \label{fig:jensens_plots}
\end{figure*}

We will first look at the change in the distribution $p_{\theta}(\lambda)$ at $\theta=\theta'$ after some small Markovian noise has been applied for a time $\Delta t$. For Markovian noise describable by a Lindblad map $\mathcal L$ with jump operator $\sqrt{\gamma}\hat L$ we may write the probability distribution $p_{\theta'}(\lambda,\Delta t)$ of $\hat \rho_{\theta'}$ using the POVM $\{\hat E_\lambda\}$, to first order in $\Delta t$, as 
\begin{align}
    p_{\theta'}(\lambda,\Delta t)&=\Tr(\hat E_\lambda(\hat \rho_{\theta'}+\Delta t\mathcal L\hat \rho_{\theta'})).
\end{align}
Further, if $\hat \rho_{\theta'}$ is pure and $\Tr(\hat E_{\lambda'}\hat \rho_{\theta'})=0$ at $\lambda=\lambda'$, then using the lemma in App.~\ref{app:trace_proof} we may write
\begin{align}
    p_{\theta'}(\lambda',\Delta t)=\gamma\Delta t\Tr(\hat E_{\lambda'} \hat L\hat \rho_{\theta'} \hat L^\dagger).\label{eq:added_support}
\end{align}
If the above quantity is nonzero, then the noise adds $\lambda'$ to the support $\mathcal S_{\theta'}$. Since the support of the noiseless probability distribution does not contain $\lambda'$ (causing a discontinuity), when Eq.~(\ref{eq:added_support}) is nonzero, this discontinuity has been removed. Formally we may state that if the initial pure state distribution has $p_{\theta'}(\lambda')=0$ and $\Tr(\hat E_{\lambda'} \hat H\hat \rho_{\theta'} \hat H)>0$ (for generator $\hat H$) then that discontinuity can be removed by small application of the noise channel $\mathcal L$ if $\Tr(\hat E_{\lambda'}\hat L\hat \rho_{\theta'} \hat L^\dagger)>0$. The removal of this discontinuity implies that continuity is restored to the Fisher information as a function of $\beta$. The restoration of continuity implies that Fisher information will tend to decrease significantly near discontinuities, at least when the noise is expected to generally \emph{decrease} the classical Fisher information \footnote{One could always construct noise models, especially nonunital ones, which \emph{increase} Fisher information. However, in many practical situations one expects the FI to decrease, and thus noise restoring continuity implies that the FI near the discontinuity must decrease to approximately the value of the original discontinuity.}.

The condition derived above has an interesting consequence. If a jump operator $\hat L$ \emph{does not} restore continuity, then the Fisher information of nearby measurements need not decrease to the value of the discontinuity in the presence of noise. To demonstrate this we will construct a pathological example below which intentionally does not restore continuity for a particular discontinuity. We can define an operator $\hat L_{\lambda'}$ which obeys 
\begin{equation}
    \Tr(\hat E_{\lambda'}\hat L_{\lambda'}\hat \rho_{\theta'} \hat L_{\lambda'}^\dagger)=\Tr(\hat E_{\lambda'}\hat H\hat \rho_{\theta'} \hat H^\dagger)=0\label{eq:pathological_condition}
\end{equation}
and thus $\hat L_{\lambda'}$ does not change the support of the probability distribution. 

In Fig.~\ref{fig:removed_discontinuities} we explicitly construct operators $\hat L_\lambda$ for the case studied in Sec.~\ref{sec:examples}. In that example, each discontinuity corresponded to a measurement outcome $\lambda=M$ having zero probability ($M$ being the outcomes of the observable $\hat J_z$). The initial probe state is a Dicke state $\ket\psi=\ket{N/2,N/2-1}$, and the the parameter $\theta$ is encoded via a unitary rotation about the $\hat J_y$ axis. The value of $\beta$ which leads to $\ket\psi$ being orthogonal to outcome $M$ is given by $\beta_{M}=\arccos(M/J)$ (see App.~\ref{app:dicke_state_analysis}). Thus Eq.~(\ref{eq:pathological_condition}) gives us the condition
\begin{equation}
    \mel{J,M}{\hat U_{\beta_{M}}^\dagger \hat L_{M}}{\psi}=0.\label{eq:disc_special_case}
\end{equation}
One can then find specific operators $\hat L_{M}$ satisfying this condition (a method of constructing them is given in App.~\ref{sec:explicit_construction}), and examine the effect these jump operators have on the Fisher information as a function of $\beta$. In Fig.~\ref{fig:removed_discontinuities}(a,b,c) the values of $M=2,8,14$ are chosen (respectively) as examples. We note that discontinuities associated with these values of $M$ remain, but the Fisher information near these discontinuities is unaffected, resulting from the fact that $\hat L_{M}$ was chosen such that it does not restore continuity.

\begin{figure*}[t]
    \centering
    \includegraphics[width=\linewidth]{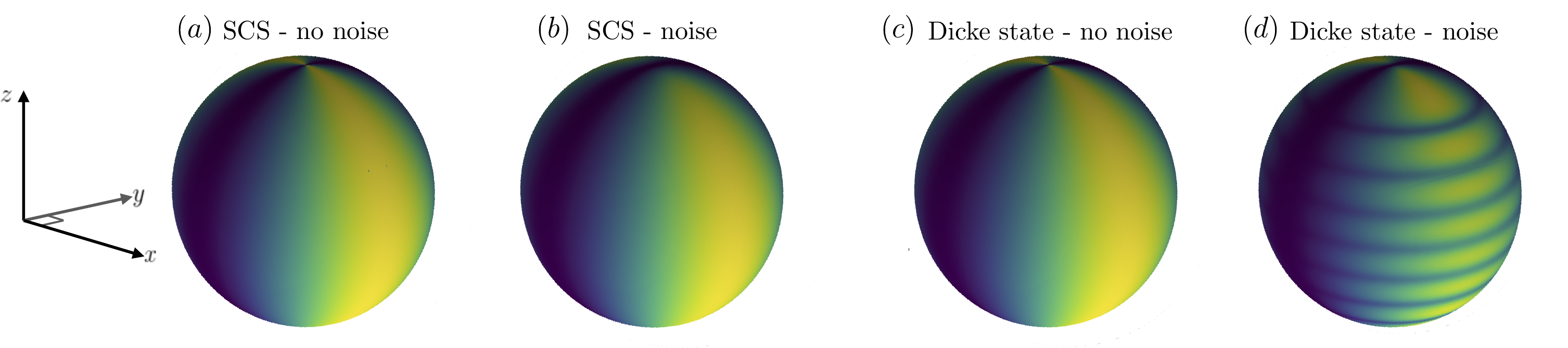}
    \caption{Classical Fisher information for a spin coherent state (SCS) pointing along $(\theta,\phi)$ (a--b) and for a first excited Dicke state (with spin $J=4$) pointing along $(\theta,\phi)$ (c--d). In both cases the brightest yellow represents regions where the classical Fisher information is equal to the quantum Fisher information of the noiseless case, and the darkest blue is CFI=0. The CFI is measured along the $\hat J_z$ axis, and the generator of $\theta$ is $\hat J_y$. The noise considered here is mixing with the identity, where the noisy state $\hat\rho$ is related to the pure state $\ket\psi$ by $\rho=(1-\varepsilon)\ketbra{\psi}+\varepsilon\mathds1/d$ where $\varepsilon=0.01$ and $d$ is the dimension of the Hilbert space.}
    \label{fig:sphere_plots}
\end{figure*}

\subsection{General Mixed States}
\label{sec:general_noise}

Both loss of signal and restoration of continuity are valid ways to interpret the decrease in Fisher information due to noise near discontinuities. We seek to generalize this, irrespective of the noise model. To simplify the analysis, in the following work we will perform a change of frame from fragile measurements to fragile probe states. Previously we considered applying a unitary rotation to the measurement basis parameterized by $\beta$ prior to measurement. Equivalently, we can fix the measurement basis to the fiducial case, and apply the inverse unitary to the noisy probe state $\hat\rho_\theta \rightarrow \hat\rho_{\theta,\beta}$.  We will refer to pure probe states $\ket{\phi}$ for which $\braket{\lambda}{\phi}=0$ for at least a single $\lambda$ as \textit{fragile probe states}. When the unitary encoding $\hat U_\theta$ is applied to these we will write them as $\ket{\phi_\theta}=\hat U_\theta \ket{\phi}$.  

We begin by recalling that the Fisher information is a convex function in the probability distribution, and thus satisfies Jensen's inequality, 
\begin{equation}
    \mathcal F_C[p_\theta(\Lambda);\theta]\leq\sum_iw_i\mathcal F_C[p_\theta^{(i)}(\Lambda);\theta]
\end{equation}
where $p_\theta(\lambda)=\sum_iw_ip_\theta^{(i)}(\lambda)$ and $\{w_i,p_\theta^{(i)}\}$ is an ensemble of distributions with nonnegative weights $w_i$ which sum to 1. When applied to mixed states this implies that the CFI is convex with respect to statistical mixtures of density operators,
\begin{align}
    \mathcal F_C[\hat\rho_\theta,\mathcal M]\leq\min_{\{w_k,\medket{\psi_\theta^{(k)}}\}}\sum_iw_i\mathcal F_C[\medketbrasingle{\psi_\theta^{(i)}},\mathcal M],\label{eq:jensen_states}
\end{align}
where $\mathcal M$ is a POVM, $\hat\rho_\theta=\sum_iw_i\medketbrasingle{\psi_\theta^{(i)}}$, and the minimum is taken over all ensembles decompositions of $\hat\rho_\theta$ into pure states. This means that the classical Fisher information of a mixed state $\hat\rho_\theta$ is upper bounded by the ``average'' of the CFIs of each pure state in a particular ensemble decomposition of $\hat\rho_\theta$.


Consider, thus, a family of mixed probe states, $\{\hat{\rho}_{\theta,\beta}\}$ for various $\beta$, in the presence of noise. To understand the effect of decoherence on the CFI we write an ensemble decomposition,
\begin{equation}
    \hat\rho_{\theta,\beta}=\sum_iq_{i,\beta}\medketbrasingle{\phi^{(i)}_{\theta }}+\hat\Omega_{\theta,\beta},\label{eq:bad_decomp}
\end{equation}
where $\{\medket{\phi^{(i)}_{\theta}}\}$ is a set of fragile pure states, and where $\hat\Omega_{\theta,\beta}\geq0$ which we refer to as the residual component. We seek the ensemble that maximizes the trace of $\sum_iq_{i,\beta}\medketbrasingle{\phi^{(i)}_{\theta }}$.  When this maximization in performed, Eq.~(\ref{eq:bad_decomp}) represents a choice of ensemble decomposition which one would expect to come close to minimizing Eq.~(\ref{eq:jensen_states}) when the noisy probe has high overlap with fragile states, as a fragile state corresponds to a discontinuous drop in the CFI. When $\medmel{\phi^{(i)}_{\theta}}{\hat\rho_{\theta,\beta}}{\phi^{(i)}_{\theta}}\approx1$, for some fragile state, one expects this particular decomposition to make Eq.~(\ref{eq:jensen_states}) especially tight.

\begin{figure*}
    \centering
    \includegraphics[width=\linewidth]{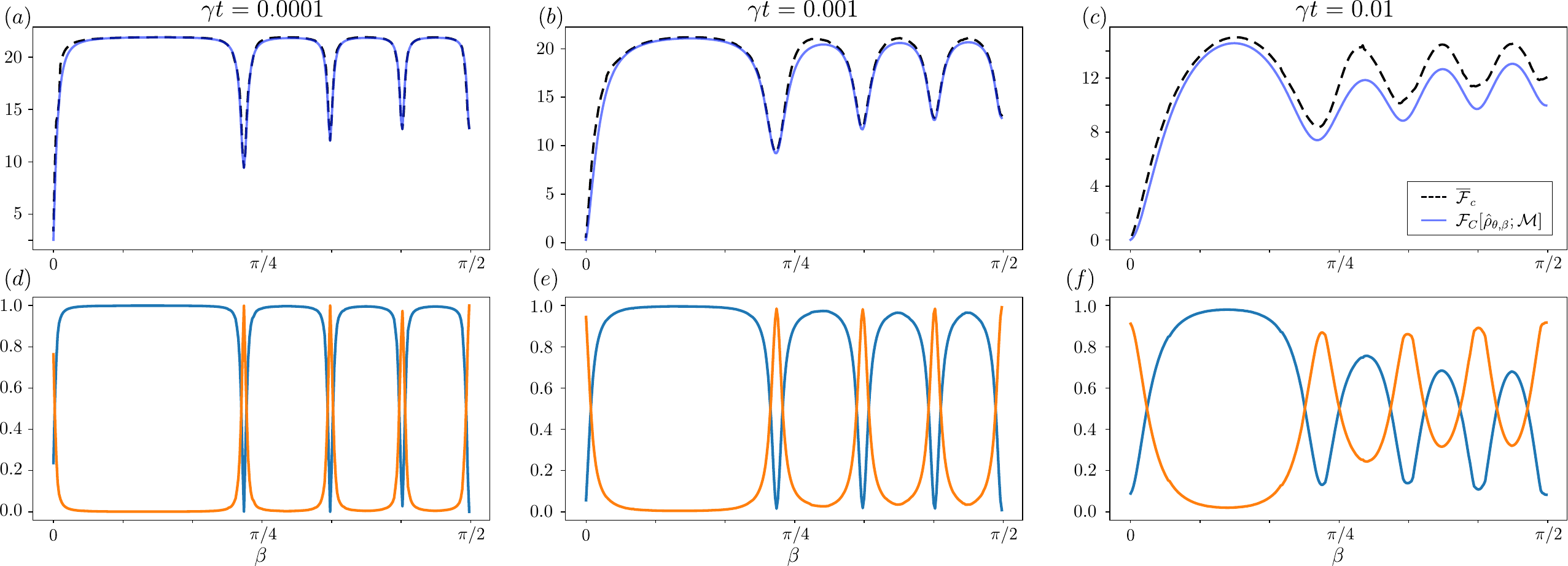}
    \caption{Top row (a-c): Classical Fisher information (blue) of the probe state corresponding a first excited Dicke state with $J=4$, after being subjected the noise channel, Eq.~(\ref{eq:loc_noise}). The probe state is rotated about the $\hat J_y$ axis (after encoding of $\theta$) by angle $-\beta$ before being measured along the $\hat J_z$ axis. The dashed line is the Jensen's upper bound, Eq.~(\ref{eq:avg_fc}), determined from the pure state decomposition in Eq.~(\ref{eq:bad_decomp}). For weak noise, the bound is tight. Bottom row (d-f): traces of $\sum_iq_{i,\beta}\medketbrasingle{\phi_\theta^{(i)}}$ (orange) and $\hat \Omega_{\theta,\beta}$ (blue) of the decompositions in (a-c).}
    \label{fig:local_noise_jensen}
\end{figure*}

To test whether the decomposition, Eq.~(\ref{eq:bad_decomp}), truly provides a tight upper bound on the CFI, consider the average CFI of the members of the ensemble, 
\begin{equation}
   \overline{\mathcal{F}}_c\left[ \hat{\rho}_{\theta,\beta}, \mathcal M \right] \equiv\sum_iq_{i,\beta}\mathcal F_C[\medketbrasingle{\phi^{(i)}_{\theta}},\mathcal M]+\mathcal F_C[\hat\Omega_{\theta,\beta},\mathcal M].\label{eq:avg_fc}
\end{equation}
In the upper row of Fig. \ref{fig:jensens_plots} we show the CFI for a first excited Dicke state of $N=8$ spins compared to $\overline{\mathcal{F}}_c$, for $\theta=0$. We subject the probe state to some amount of collective depolarizing noise described by Eq.~(\ref{eq:collective_lindbladian}) prior to sensing. We consider the state rotated around the $y$-axis by $\beta$ and measured in the fiducial basis $\ket{J,M}$ with $J=4$. We find that $\overline{\mathcal{F}}_c$ in Fig.~\ref{fig:jensens_plots} comes near to saturating the CFI when the rotated probe state is near to a fragile state, demonstrating that the decomposition in Eq.~(\ref{eq:bad_decomp}) is near to the true minimizing decomposition. In the bottom row of Fig.~\ref{fig:jensens_plots} we also provide the traces of both the residual component $\hat\Omega_{\theta,\beta}$ and the component in the span of fragile states. This highlights that in general one does not expect an arbitrary state to have an ensemble decomposition consisting only of fragile states. To see this we note that the trace of $\sum_iq_{i,\beta}\medketbrasingle{\phi_\theta^{(i)}}$ is near 0 when $\beta$ is far from discontinuities in Fig.~\ref{fig:jensens_plots}.

We interpret this as meaning that when one measures using a probe state very near to a fragile one, the presence of noise can enable one to write a decomposition which predominantly contains fragile states $\medket{\phi_\theta^{(i)}}$ for which the CFI is discontinuous. This necessarily lowers $\overline{\mathcal F}_c$, and thus sets a tight upper bound on the CFI of the noisy probe state.

Given the tightness of the plots in Fig.~\ref{fig:jensens_plots} for early times, one may wonder if the ``minimizing decomposition'' (that which makes $\overline{\mathcal F}_c$ equal to the CFI of the noisy probe state) contains fragile states. Surprisingly, one can show that minimizing decomposition generally cannot contains these states. To prove this, let us first consider an ensemble decomposition $\{w_i,\medketbrasingle{\psi_\theta^{(i)}}\}$ of $\hat\rho_\theta$ which saturates Jensen's inequality by assumption, i.e.,
\begin{equation}
    \mathcal F[\hat\rho_\theta,\mathcal M]=\sum_iw_i\mathcal F[\medketbrasingle{\psi_\theta^{(i)}},\mathcal M].\label{eq:sat_jensen}
\end{equation}
In general $\medketbrasingle{\psi_\theta^{(i)}}$ and $\hat\rho_\theta$ are also functions of $\beta$, but for ease of notation we leave that as an implicit assumption in the following. If Eq.~(\ref{eq:sat_jensen}) is true, then it is also true that for any $\medketbrasingle{\psi_\theta^{(j)}}$ one can write
\begin{align}
    \nonumber&\mathcal F[\hat\rho_\theta,\mathcal M]\\
    &=\mathcal F[\hat\rho_\theta-w_j\medketbrasingle{\psi_\theta^{(j)}},\mathcal M]+w_j\mathcal F[\medketbrasingle{\psi_\theta^{(j)}},\mathcal M].
\end{align}
This implies that $\mathcal F[\hat\rho_\theta-w_j\medketbrasingle{\psi_\theta^{(j)}},\mathcal M]$ is linear in $w_j$, meaning all higher derivatives than the first with respect to $w_j$ must be zero. Using this, we write that for $p_\theta(\lambda)=\Tr(\hat\rho_\theta \hat E_\lambda)$ and $g_\theta(\lambda)=\Tr(\medketbrasingle{\psi_\theta^{(j)}}\hat E_\lambda)$ we have
\begin{align}
    \nonumber&\frac{\mathrm d^{k}}{\mathrm d^kw_j}\mathcal F[\hat\rho_\theta-w_j\hat\sigma_\theta^{(j)},\mathcal M]\\
    &=\frac{\mathrm d^{k}}{\mathrm d^kw_j}\sum_\lambda \frac{(\partial_\theta p_\theta(\lambda)-w_j \partial_\theta g_\theta(\lambda))^2}{p_\theta(\lambda)-w_j g_\theta(\lambda)}=0
\end{align}
for all $k\geq2$. This condition is satisfied if and only if (see App.~\ref{sec:min_ens_dec})
\begin{equation}
    g_\theta(\lambda)\partial_\theta p_\theta(\lambda)=p_\theta(\lambda)\partial_\theta g_\theta(\lambda) \;\;\forall \;\lambda \in \mathcal S.\label{eq:min_decomp_cond}
\end{equation}
If both $p_\theta(\lambda)$ and $g_\theta(\lambda)$ share the same support in $\lambda$, then Eq.~(\ref{eq:min_decomp_cond}) enforces that the score functions of $p_\theta(\lambda)$ and $g_\theta(\lambda)$ are equal on their shared support. If either $p_\theta(\lambda)$ or $g_\theta(\lambda)$ is zero on some subset of outcomes $\lambda'\in\mathcal N_\theta$ then those distributions also have $\partial_\theta p_\theta(\lambda')=0$ and $\partial_\theta g_\theta(\lambda')=0$ respectively for unitary encodings of $\theta$. This is a natural consequence of unitary dynamics in quantum physics. Thus, for $\lambda'\in\mathcal N_\theta$ Eq.~(\ref{eq:min_decomp_cond}) is satisfied trivially.

What this means is that decomposition in Eq.~(\ref{eq:bad_decomp}) is generally not be saturable for $q_{i,\beta}>0$, as the score functions of the fragile probe states need not generally be equal to the score of the noisy probe $\hat\rho_{\theta,\beta}$. Despite this, in numerical studies we find that for moderate noise Eq.~(\ref{eq:bad_decomp}) is fairly tight, especially for $\hat\rho_{\theta,\beta}$ near fragile probe states.

This understanding becomes particularly useful when studying more general families of probe states, such as the family $\ket{J,M_\mathbf{n}}$, with projection $M_\bold{n}$ of angular momentum along direction $\mathbf{n}$ on the sphere. Figure~\ref{fig:sphere_plots} shows the classical Fisher information associated with measuring a rotation about the $y$-axis at $\theta=0$, with projective $\hat J_z$ measurements. Figure~\ref{fig:sphere_plots}(a,b) shows the CFI of a spin coherent state $\ket{J,J_\mathbf{n}}$ over the unit sphere direction $\mathbf{n}$ both for no noise, and for small depolarizing noise; Figure~\ref{fig:sphere_plots}(c,d) is equivalent for the first excited Dicke state  $\ket{J,(J-1)_\mathbf{n}}$. We observe that the spin coherent state is robust, but the first Dicke state has fragility along at certain polar angles, giving rise to the ``stripes'' seen when noise is introduced.  These stripes form near the directions $\mathbf{n}$ which cause a discontinuity and are azimuthally symmetric as rotation around the $z$-axis does not change the support of the $\hat J_z$ measurement probability distribution.


\subsection{Local noise and fragile probe states}


In contrast to the noise channel in Eq.~(\ref{eq:collective_lindbladian}), local depolarizing noise is given by the Lindbladian
\begin{equation}
    \mathcal L\hat\rho=\gamma\sum_{\nu,i}\hat\sigma_\nu^{(i)}\hat\rho\hat\sigma_\nu^{(i)}-3N\gamma\hat\rho, \label{eq:loc_noise}
\end{equation}
where $\sigma_\nu^{(i)}$, with $\nu=\{x,y,z\}$, are the Pauli operators acting locally on the $i^{th}$ spin. To describe the effects this noise, we utilize the collective space formalism described in \cite{Forbes2024,Chase2008} and model the system using angular momentum eigenstates in irreducible representations of SU(2) with total angular momentum $J<N/2$. Using this description we can perform the same decomposition as Eq.~(\ref{eq:bad_decomp}), but with fragile states $\medket{\phi_\theta^{(i)}}$ in each irrep labeled by $0\leq J\leq N/2$. The fragile probe states in each irrep can be found using the results in App.~\ref{app:dicke_state_analysis}.

We observe in Fig.~\ref{fig:local_noise_jensen} that, just as with collective noise, decomposing the state by maximizing its overlap with fragile pure states, Eq.~(\ref{eq:bad_decomp}), sets a fairly tight upper bound for moderate amounts of noise. In Fig.~\ref{fig:local_noise_jensen}, the upper bound, represented by the dashed line, approximately saturates the true CFI for times around $\gamma t=0.001$ and earlier. At later times, the noise is so large that this decomposition strays more from saturating the true CFI, as seen at time $\gamma t=0.01$. This is also observed in the collective noise case in Fig.~\ref{fig:jensens_plots}.

\begin{figure}
    \centering
    \includegraphics[width=1\linewidth]{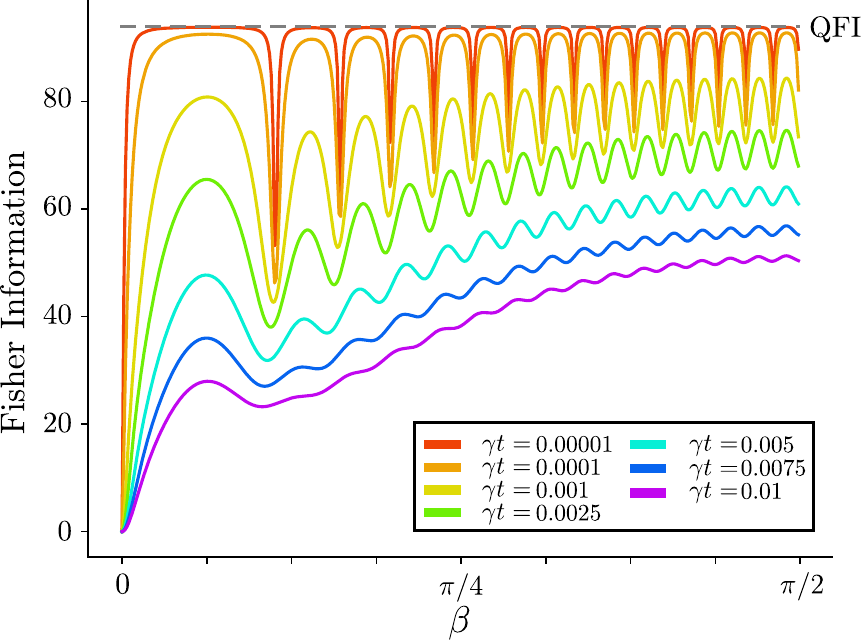}
    \caption{Classical Fisher information for a $J=16$ spin first excited Dicke state, undergoing local depolarizing noise Eq.~(\ref{eq:loc_noise}) at a rate $\gamma$. The Fisher information is with respect to sensing a rotation $\theta$ about the $\hat J_y$ axis and measuring in the $\hat J_z$ eigenstate basis. The values of $\gamma t$ shown here are the same as in Fig.~\ref{fig:w_state} to demonstrate how equal rates of collective noise vs local noise affect the Fisher information differently.}
    \label{fig:32_qubit_local_noise}
\end{figure}

In Fig.~\ref{fig:32_qubit_local_noise} we plot the CFI associated with the same protocol as Sec.~\ref{sec:examples}, but with the local noise from Eq.~(\ref{eq:loc_noise}) rather than the collective noise in Eq.~(\ref{eq:collective_lindbladian}). The CFI is displayed for different noise strengths, labeled by applying the noise channel for different amounts of time $\gamma t$. The noise strengths in Fig.~\ref{fig:32_qubit_local_noise} are the same as in Fig.~\ref{fig:w_state} to demonstrate the difference in effects of local noise versus collective noise on the CFI. We note that local noise appears to be less damaging to the CFI in the $\beta\approx\pi/2$ regime than collective noise.

\subsection{Approximate Loss of Fisher Information}\label{sec:robust_measurements}


We will approximate the difference between the Fisher information of some initial distribution $p_\theta(\lambda)$ and one which is subject to a small amount of noise $P_\theta(\lambda,\Delta t)=p_\theta(\lambda)+\gamma\Delta t\sigma_\theta(\lambda)$ up to first order in $\gamma\Delta t$, where
\begin{align}
    \sum_\lambda\sigma_\theta(\lambda)=0.
\end{align}
For this distribution we write that the change in the CFI over a small time $\gamma\Delta t$ is
\begin{align}
    \nonumber\mathcal F_C[P_\theta(\Lambda,0);\theta]&-\mathcal F_C[P_\theta(\Lambda,\Delta t);\theta]\\
    &\approx\sum_\lambda\frac{\Delta\mathcal F_C^{(\lambda)}[p_\theta(\Lambda);\theta]}{\frac1{\gamma\Delta t\sigma_\theta(\lambda)}p_\theta(\lambda)+1},
    \label{eq:cfi_appx}
\end{align}
where $\Delta\mathcal F_C^{(\lambda)}[p_\theta(\Lambda);\theta]$ is the size of the discontinuity which occurs when $\lambda\in\mathcal N_\theta$. Details of this approximation can be found in App.~\ref{sec:appx_loss_details}. One may interpret Eq.~(\ref{eq:cfi_appx}) as saying that for small $\gamma\Delta t$ the CFI decreases by $\Delta \mathcal F_C^{(\lambda)}[p_\theta(\Lambda);\theta]$ when $p_\theta(\lambda)\approx0$. However, as $p_\theta(\lambda)$ increases, we expect to see an increase in the CFI determined by the overlap between the initial pure state and the POVM element $\Tr(\hat\rho_\theta \hat E_\lambda)=p_\theta(\lambda)$. This is precisely what is seen in Fig.~\ref{fig:w_state}, where the decrease in the Fisher information is approximately equal to the discontinuity near points where $p_\theta(\lambda')\approx0$ for some $\lambda'$, but the loss of CFI decays as $\beta$ is swept to either side of the discontinuity. We may further glean from Eq.~(\ref{eq:cfi_appx}) that the width $\Gamma$ of the ``wells'' in plots like figures \ref{fig:w_state} and \ref{fig:32_qubit_local_noise} scales as $\Gamma\sim\sqrt{\gamma\Delta t}$ for short times.

Since $p_\theta(\lambda)$ is simply the trace inner product between $\hat\rho_\theta$ and $\hat E_\lambda$ we may interpret $p_\theta(\lambda)$ as measuring the ``closeness'' between $\hat\rho_\theta$ and the space orthogonal to $\hat E_\lambda$, with projector onto the space $\hat{\mathds P}_\lambda^\perp$. When $p_\theta(\lambda)$ is 0, the state $\hat\rho_\theta$ is maximally close to $\hat{\mathds P}_\lambda^\perp$, and when $p_\theta(\lambda)=1$ the state $\hat\rho_\theta$ is maximally far from $\hat{\mathds P}_\lambda^\perp$. This is not a proper notion of distance; however, it is convenient to think about it as such when interpreting Eq.~(\ref{eq:cfi_appx}). Using this, we note that fragile states lie in this projector space $\medketbrasingle{\phi_\theta^{(i)}}\in\hat{\mathds P}_\lambda^\perp$ for $\lambda\in\mathcal S$, and thus $p_\theta(\lambda)\gg0$ implies that $\hat\rho_\theta$ has low overlap with fragile states.

Finally, we note that the approximation in Eq.~(\ref{eq:cfi_appx}) relies on small $p_\theta(\lambda)\sim\gamma \Delta t$, which cannot be satisfied simultaneously for all $\lambda\in\mathcal S$. However, when $p_\theta(\lambda)$ does not satisfy this condition, the summand in Eq.~(\ref{eq:cfi_appx}) is itself on the order of $\gamma \Delta t$ to first order in $\gamma\Delta t$. Thus the approximation in Eq.~(\ref{eq:cfi_appx}) is accurate up to these terms of order $\gamma\Delta t$.


\section{Discussion}
\label{sec:discussion}

In the following we will address additional considerations regarding the behavior of discontinuities and their effects on the CFI in various scenarios. We discuss here discontinuities in the large $N$ limit, the behavior of the maximum likelihood estimator near discontinuities from the example in Sec.~\ref{sec:examples}, and an example of discontinuities in protocols from the literature.

\subsection{Large $N$ limit}

From the example seen in Sec.~\ref{sec:examples}, one might be concerned that in the limit as $N\to\infty$ measurements where $\beta\approx\pi/2$ are fragile and potentially non-optimal. This would present a contradiction with known results from sensing using a bosonic mode, as the limit as $N\to\infty$ can be approximated by a bosonic mode in the Holstein-Primakoff approximation (HPA). More specifically, in the HPA one can map the first excited Dicke state (along $\hat J_z$ in this case) to a Fock state with $n=1$ excitation. Rotations are mapped to displacement, and measurements in the eigenbasis of linear combinations of $\hat J_x$ and $\hat J_y$ are mapped to homodyne measurement (measurement in the eigenbasis of the position and momentum operators $\hat X$ and $\hat P$). One can show that homodyne measurement is optimal for measuring displacements of Fock states, and so one would expect that optimally measuring rotations about $\hat J_y$ using Dicke states along the $\hat J_z$ direction would be achieved by measuring in the eigenbasis of $\hat J_x$. In order for this to be true, the size of the discontinuities relative to the total Fisher information for $\beta$ near $\pi/2$ should tend to 0 as $N\to\infty$. We show in App.~\ref{app:dicke_state_analysis} that the ratio between the size of discontinuities near $\beta\approx\pi/2$ over the total Fisher information scales as
\begin{equation}
    \frac{\Delta \mathcal F_C[\hat\rho_\theta;\text{POVM}_{\beta\approx\pi/2}]}{\mathcal F_C[\hat\rho_\theta;\text{POVM}_{\beta\approx\pi/2}]}\sim\frac{1}{\sqrt{J}},
\end{equation}
for large $J$. Thus, in the regime where HPA is valid, measuring in the eigenbasis of linear combinations of $\hat J_x$ and $\hat J_y$ is asymptotically optimal.

\begin{figure}
    \centering
    \includegraphics[width=\linewidth]{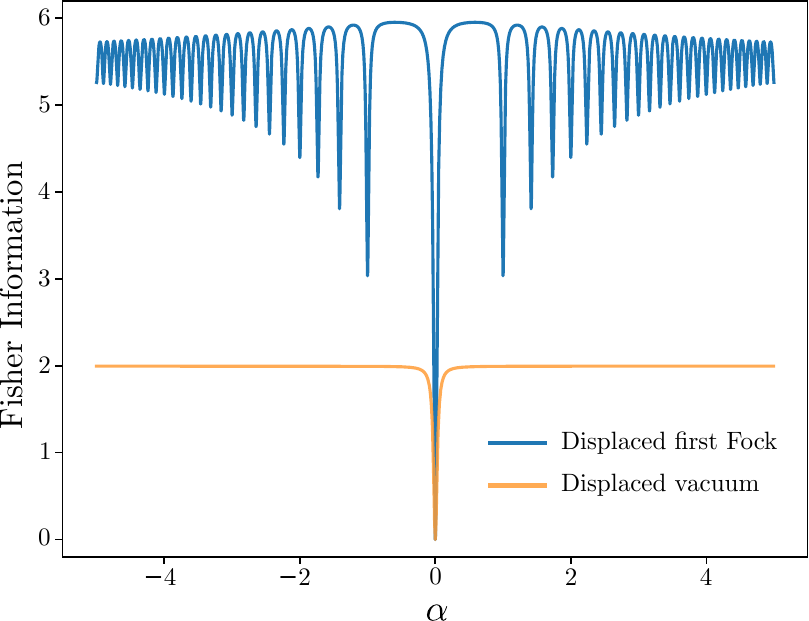}
    \caption{Fisher information of a noisy displaced vacuum and noisy displaced first Fock state as a function of displacement $\hat D(\alpha)$. The noise model is a Lindblad channel with jump operators $\sqrt\gamma \hat X$ and $\sqrt\gamma\hat P$ applied over time $\gamma t=0.001$ prior to sensing. The parameter $\theta$ is encoded by displacement operator $e^{-i\theta \hat P}$ and the Fisher information is calculated by measuring in the number basis $\{\ketbra n\}$.}
    \label{fig:bosonic_cfi}
\end{figure}

We also show in App.~\ref{app:dicke_state_analysis} that for $\beta\sim 1/\sqrt J$, the size of the discontinuity scales as $\sim J$, meaning the size of discontinuities remains on the same order as the total Fisher information for these measurements. If we take $\beta\approx0$, then this corresponds to states in the HPA being measured using the number basis eigenstates $\ket n$. As such, we should expect to observe discontinuities in the bosonic mode when sensing displacements using a displaced first Fock state by measuring in the number basis. We observe this in Fig.~\ref{fig:bosonic_cfi} where we plot the CFI for the first Fock state, and we observe discontinuities as a function of the displacement $\alpha$. The parameter $\theta$ is encoded by the generator $\hat P$, and the CFI is evaluated at $\theta=0$. The discontinuities seen in Fig.~\ref{fig:bosonic_cfi} can be thought of as the large limit of $J$ for sensing a rotation of a first Dicke state and measuring in the $\hat J_z$ basis.

\subsection{Maximum Likelihood estimator near discontinuities}

Given that discontinuities at $\beta\sim 1/\sqrt J$ do not vanish as $J\to\infty$, does sensing in this regime ever present an advantage? Consider, for example, the case $J=16$ shown in Fig.~\ref{fig:w_state}. The point at which CFI is largest, in the presence of noise, is $\beta\approx0.2$, which lies in the small $\beta\sim J^{-1/2}$ regime. 

\begin{figure}[t]
    \centering
    \includegraphics[width=\linewidth]{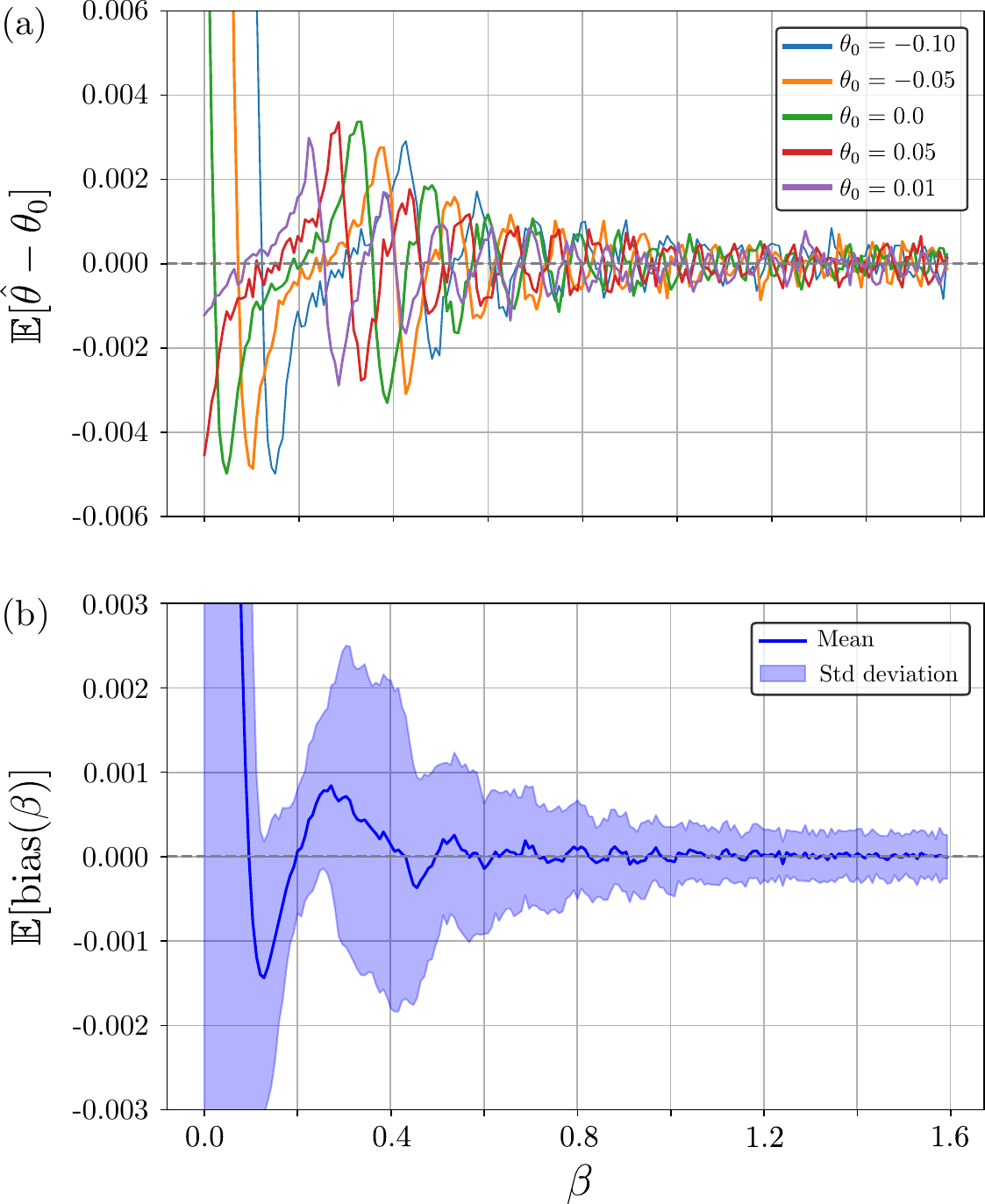}
    \caption{Plot of the biases of the maximum likelihood estimator $\hat\theta$ as a function of $\beta$ from 0 to $\pi/2$ for a noisy first excited Dicke state ($J=16$) rotated by $\beta$ relative to the $\hat J_z$ direction. The expectation value is calculated using 40 samples, and averaged over 10,000 runs. The probe state has undergone the noise channel in Eq.~(\ref{eq:collective_lindbladian}) up to $\gamma t=0.01$. In (a), multiple values of the true values $\theta_0$ are plotted, and in (b) an average is taken over $\theta_0$ on the range of $[-0.1,0.1]$. We observe that the average bias overall increases for smaller $\beta$ on this averaged range of $\theta_0$, indicating that measuring in this regime could lead to large bias. Maximum likelihood estimation was performed over the range $[-0.2,0.2]$ for $\theta_0$, and indistinguishability near $\beta=0$ was removed by enforcing that negative values of $\hat\theta$ are mapped to their positive counterparts.}
    \label{fig:bias_plot}
\end{figure}

While the CFI determines the ultimate sensing precision according to the Cram\'er-Rao bound in the asymptotic limit of $n\rightarrow \infty$ samples, in practice, the actual uncertainty in the parameter will depend on the nature of the estimator for a finite sample size.  An important consideration is the bias of the maximum likelihood estimator, which we will examine here. The bias is the difference between the expected value of the estimator and the true parameter, and thus captures how well one can identify the true parameter given a finite number of samples. We calculate the bias for a noisy Dicke state as a function of $\beta$ in Fig.~\ref{fig:bias_plot}. Here, the probe state is subjected to the noise channel in Eq.~(\ref{eq:collective_lindbladian}) up to a time $\gamma t=0.001$. This time was chosen since it exhibits the behavior that the CFI near $\beta\approx0.2$ is larger than the CFI near $\beta\approx\pi/2$, as can be seen in Fig.~\ref{fig:w_state}. In Fig.~\ref{fig:bias_plot}a the expected value of the bias is calculated for 40 samples over 10,000 runs with multiple true values of the parameter $\theta_0$ shown. In Fig.~\ref{fig:bias_plot}b, an average over possible values of $\theta_0$ is taken, averaged uniformly over the range $\theta_0=[-0.1,0.1]$. We find that while there are certain values of $\beta$ and $\theta_0$ for which the bias is zero near $\beta\approx0.2$, this point is a function of $\theta_0$ itself. Further, the bias steeply changes to either side of the point where $\mathbb E[\hat\theta-\theta_0]=0$. 

The deviation shown in Fig.~\ref{fig:bias_plot}b highlights that uncertainty in $\theta_0$ can cause the average bias to increases substantially in the $\beta\approx0.2$ region, whereas the bias in the $\beta\approx\pi/2$ region appears to be only a weak function of $\theta_0$. This implies that a priori knowledge of the parameter $\theta_0$ may allow one to construct a very precise measurement in the $\beta\approx0.2$ regime (with low bias), but for more general sensing tasks where the value of $\theta_0$ is not known a priori to some fine precision, the more standard measurement regime of $\beta\approx\pi/2$ is most appropriate.

\subsection{Discontinuities in Loschmidt Echo Protocols}

Finally, we would like to point out that discontinuities can appear in a well-known method of optimal quantum measurement based on the Loschmidt echo \cite{Pezze2016} (also referred to as time reversal \cite{Colombo2022}). In such protocols, one often begins with an uncorrelated quantum state, followed by a highly nonlinear entangling Hamiltonian. This serves as the probe state for sensing a linear encoding of $\theta$. After encoding, one applies the inverse of the entangling interaction, followed by measurement. Such protocols have demonstrated robustness to noise \cite{Davis2016,Frowis2016}, as well as amplified measurement signal when compared to their non-entangled counterparts \cite{Colombo2022,Li2023}. However, how one measures the final state in these protocols is of critical importance to the robustness of the measurement to noise in the probe, encoding, and measurement device. Allowing a POVM element to be the projector onto the initial unentangled state, as proposed in \cite{Pezze2016}, while saturating the quantum Fisher information, potentially leads to a fragile measurement for small $\theta$.

For a pure probe state $\ket\psi$ with parameter $\theta$ encoded by Hamiltonian $\hat H$ one can always write the state after encoding of $\theta$ as
\begin{equation}
    e^{-i\theta\hat H}\ket\psi=\alpha\ket\psi+\beta\ket{\psi_\perp},
\end{equation}
where $\braket{\psi}{\psi_\perp}=0$ and $|\alpha|^2+|\beta|^2=1$. In such cases, the state can be optimally measured in the basis of 
\begin{equation}
    \{\ketbra\psi,\ketbra{\psi_\perp}\}.\label{eq:same_or_not}
\end{equation}
However, when $\theta=0$, this basis experiences a discontinuity since the probability of observing $\ket{\psi_\perp}$ is 0. Thus, despite technically forming an optimal measurement basis for pure states, this particular basis will typically be fragile to noise. Protocols involving time-reversal have been experimentally implemented \cite{Linnemann2016} and demonstrate robustness when one measures in a way which does not include projection onto the initial state.


\section{Summary and Outlook}

The classical Fisher information, unlike the quantum Fisher information, is subject to discontinuities under unitary rotation of the initial probe state. In this work, we demonstrated a connection between states whose Fisher information decreases rapidly under noise and proximity to ``fragile states'', for which the classical Fisher information experiences a discontinuity. This discontinuity is a result of one or more outcomes occurring with probability zero, thus leading to a lack of signal on that outcome. We use this interpretation of proximity to fragile states to find pure state decompositions of the mixed probe state which come near to saturating the upper bound set by Jensen's inequality. We show that while fragile states typically do not constitute a decomposition that exactly minimizes Jensen's inequality, at least for moderate amounts of noise, the decomposition involving fragile states saturates the true CFI approximately. We further used this interpretation to explain the difference between how collective noise and local noise affect the Fisher information. We provided an approximation to the loss of Fisher information near fragile states under Markovian noise. This approximation demonstrates how loss of Fisher information depends directly on the size of the discontinuity, as well as the overlap between the initial probe state and fragile states.

We looked at the large spin limit, and compared it to known results about optimal measurements in bosonic modes. We found that in certain regimes probe states can become asymptotically optimal in the large $J$ limit. However, for certain probe states which are very aligned with the measurement axis, Fisher information discontinuities do not disappear as $J\to\infty$. These discontinuities can then be seen in the bosonic limit when sensing displacements with a Fock state, and measuring in the number state basis.

While in this work we focused on examples with angular momentum eigenstates and Fock states, these discontinuities also occur in measurements of many commonly proposed probe states, such as N00N states, GHZ states, twin Fock states, and oversqueezed spin states. In many cases, there may be a family of measurements, related by an easily implementable unitary, which all provide maximum Fisher information for some pure probe state $\ket{\psi_\theta}$. However, these measurements may have drastically different Fisher information when noise is applied to the probe.



Further work is needed to characterize how the fragility of the Fisher information extends from discrete-outcome POVMs to POVMs with a continuum of outcomes, such as homodyne and heterodyne detection. For continuous-outcome POVMs, the discontinuities identified above need not arise.
Nevertheless, the attainable Fisher information can still vary across POVMs, and some optimal continuous measurements can strictly outperform others. Whether similar arguments in Sec.~\ref{sec:noise} can be applied to understand the fragility of the Fisher information in the continuous-outcome regime is an open question.



\section{Acknowledgments}
This work is supported by funding from the NSF Quantum Leap Challenge Institutes program, Award No. 2016244, the NSF Grant PHY-2116246, and the National Research Council of Canada.

\bibliographystyle{apsrev4-3}
\bibliography{refs.bib}

\appendix

\section{Phase Estimation of a Qubit}
\label{sec:qubit}

In this section we will study the classical Fisher information of a noisy qubit with respect to sensing rotations. We demonstrate how the CFI decreases when measuring aligned with the initial probe state, and relate this to proximity to fragile states.

Consider a qubit, prepared in the $\ket{+}=\ket{\uparrow_x}=\frac1{\sqrt2}\ket{\uparrow}+\frac1{\sqrt2}\ket{\downarrow}$ state. We wish to use this state to sense a rotation around the $z$-axis by an angle $\theta$. The state $\hat \rho_\theta $ is given by
\begin{equation}
    \hat \rho_\theta =e^{-i\theta\hat\sigma_z/2}\ketbra{\uparrow_x}e^{i\theta\hat\sigma_z/2}.
\end{equation}

It is an easy exercise to show that classical Fisher information obtained by measuring the pure state above using a POVM in the equator of the Bloch sphere (linear combinations of $\hat\sigma_x$ and $\hat\sigma_y$) is equal to the QFI, and thus constitutes an ``optimal measurement''. Explicitly, the POVM elements are given by
\begin{align}
    &\text{POVM}_{\beta,\theta}\nonumber\\&=\{e^{-i\beta\hat\sigma_z/2}\hat\rho_\theta e^{i\beta\hat\sigma_z/2},e^{-i\beta\hat\sigma_z/2}\hat\rho_{\theta+\pi}e^{i\beta\hat\sigma_z/2}\},
\end{align}
where $\beta\in[0,\pi)$, and parametrizes the angle at which we are measuring relative to $\hat \rho_\theta$. The probabilities are
\begin{align}
    p_\uparrow(\theta,\beta)=\cos^2\left(\frac{\beta-\theta}{2}\right)\\
    p_\downarrow(\theta,\beta)=\sin^2\left(\frac{\beta-\theta}{2}\right)
\end{align}
We see that if $\beta=\theta+n\pi$ for $n\in\mathds N$ there will be outcomes with probability zero, and thus a potential discontinuity. In fact, the CFI is maximum for all $\beta$ and $\theta$ except at these discrete points, for which the CFI is equal to 0.

\begin{figure}[t]
    \centering
    \includegraphics[width=\linewidth]{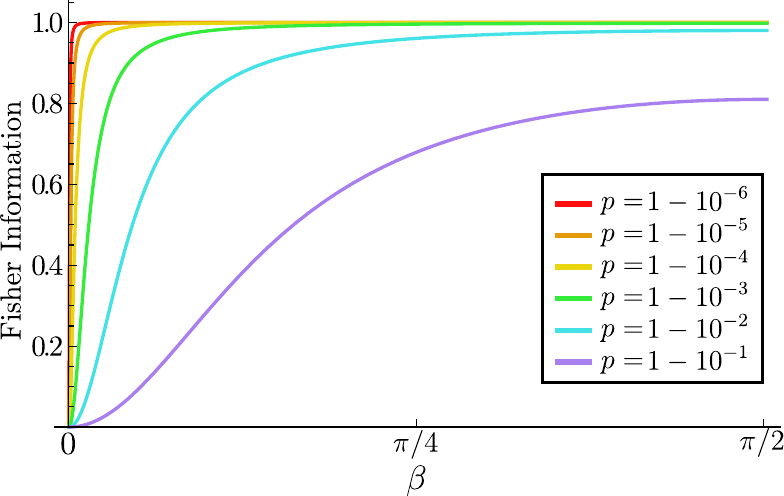}
    \caption{Classical Fisher information for different amounts of mixing with identity as a function of the measurement angle $\beta$ for $\theta=0$. The parameter $p$ is defined in Eq.~(\ref{eq:depolarized_qubit}) and parameterizes the amount of depolarizing noise. When $\beta\approx0$, we see that the CFI drops to near zero, despite the fact that in the absence of noise, such a measurement would be considered optimal.
    }
    \label{fig:qubit_cfi}
\end{figure}

Na\"ively, one may expect that any measurement in any POVM besides $\beta=\theta+n\pi$ is an equally good measurement, in the sense that they all result in the maximum possible CFI. However, we now introduce a small amount of depolarizing noise to the probe state and find that measurements near $\beta=\theta+n\pi$ have a comparable amount of Fisher information to that of the noiseless state at the point of discontinuity, as seen in Fig.~\ref{fig:qubit_cfi} for which $\theta=0$. Explicitly, we find that for small depolarizing $p$ such that the state is
\begin{equation}
    \hat \rho_\theta(p)=e^{-i\theta\hat\sigma_z/2}\left((1-p)\ketbra{\uparrow_x}+\frac p2\mathds1\right)e^{i\theta\hat\sigma_z/2}.\label{eq:depolarized_qubit}
\end{equation}
one obtains outcome probabilities
\begin{align}
    p_\uparrow(\theta,\beta)&=\frac12-\frac{p-1}{2}\cos(\beta-\theta),\\
    p_\uparrow(\theta,\beta)&=\frac12+\frac{p-1}{2}\cos(\beta-\theta),
\end{align}
which yields CFI
\begin{equation}
    \mathcal F_C[\hat\rho_\theta;\text{POVM}_{\beta,\theta}]=\frac{(1-p)^2\sin^2(\beta-\theta)}{1-(1-p)^2\cos^2(\beta-\theta)}.
\end{equation}
Despite the fact that for $p>0$ the probabilities $p_\uparrow$ and $p_\downarrow$ are never 0, we see from Fig.~\ref{fig:qubit_cfi} that near where $\beta=0$ (with $\theta=0$) the CFI decreases to 0 at the point where the discontinuity was before the noise was introduced. This highlights that even for small amounts of noise $p$, measurements which were once optimal for pure probe states can have near 0 CFI when $\beta$ is near discontinuities.

\section{Proof of zero trace with POVM element}
\label{app:trace_proof}

In this appendix we will prove that if $\hat \rho$ is pure and $\Tr(\hat \rho \hat E_\lambda )=0$, then $\Tr(\hat A\hat \rho \hat E_\lambda )=0$ for any $\hat A$. First, assume a rank-$k$ POVM element defined as
\begin{equation}
    \hat E_\lambda =\sum_{n=1}^kc_n\ketbra n
\end{equation}
where $\{\ketbra n\}$ forms an orthonormal basis. Since $\hat E_\lambda $ is positive semi-definite, $c_n>0$. Therefore $\Tr(\hat \rho \hat E_\lambda )=0$ implies that
\begin{equation}
    \mel{n}{\hat \rho}{n}=0,\;\;\;\forall n.
\end{equation}
When $\hat \rho$ is pure this also implies that $\braket{\psi}{n}=0$, $\forall n$. Therefore
\begin{equation}
    \Tr(\hat A\hat \rho \hat E_\lambda )=\sum_{n=1}^k\Tr(\hat A\ket\psi\braket{\psi}{n}\bra n)=0,
\end{equation}
thus completing the proof.

\section{Discontinuities in the First Dicke State}
\label{app:dicke_state_analysis}
In this section, we will derive a few key results for the behavior of the first excited Dicke state used for sensing in Sec.~\ref{sec:examples}. We begin by deriving the probability distribution for a rotated first excited Dicke state with total angular momentum $J$
\begin{equation}
    \hat\rho_\theta=e^{-i\theta \hat J_y}\ketbra{J,J-1}e^{+i\theta\hat J_y},
\end{equation}
measured in the basis of eigenstates of the operator
\begin{equation}
    \hat J_\beta=e^{-i\beta \hat J_y}\hat J_ze^{+i\beta\hat J_y}.
\end{equation}
The expression for a spin coherent state pointing along $(\theta,\phi)$ in the $\hat J_z$ eigenbasis is well known, and given by
\begin{equation}
    \ket{\theta,\phi}=\sum_{M=-J}^Jg_{J,M}(\theta,\phi)\ket{J,M}
\end{equation}
where
\begin{align}
    \nonumber g_{J,M}(\theta,\phi)=&\sqrt{\frac{(2J)!}{(J-M)!(J+M)!}}\\
    &\times\cos(\theta/2)^{J+M}\sin(\theta/2)^{J-M}e^{-iM\phi}.
\end{align}
One may therefore write that
\begin{equation}
    e^{-i\theta \hat J_y}\ket{J,J-1}=\frac{1}{\sqrt{\mathcal N}} \hat R_y(\theta)\hat J_-\hat R^\dagger_y(\theta)\ket{\theta,0}
\end{equation}
where $\hat R_y(\theta)=e^{-i\theta \hat J_y}$, $\ket{\theta,0}$ is a spin coherent state pointing along polar angle $\theta$ and azimuthal angle $\phi=0$, and $\mathcal N$ is a normalizing factor. We find that
\begin{equation}
    \mathcal N=\mel{J,J-1}{\hat J_+\hat J_-}{J,J-1}=2J,
\end{equation}
and
\begin{equation}
    \hat R_y(\theta)\hat J_-\hat R^\dagger_y(\theta)=\cos^2(\theta/2)\hat J_--\sin^2(\theta/2)\hat J_+-\sin(\theta)\hat J_z.
\end{equation}
Plugging these in and simplifying we find that
\begin{equation}
    e^{-i\theta \hat J_y}\ket{J,J-1}=\sum_{M=-J}^Jh_M(\theta)\ket{J,M},
\end{equation}
where
\begin{align}
    \nonumber h_M(\theta)=&\sqrt{\frac{(2J-1)!}{(J-M)!(J+M)!}}\\
    \nonumber&\times \cos(\theta/2)^{J+M-1}\sin(\theta/2)^{J-M-1}\\
    &\times\left(J\left(\cos^2(\theta/2)-\sin^2(\theta/2)\right)-M\right)\\
    =\nonumber&\sqrt{\frac{2}{J}\binom{2J}{J-M}}\left(\frac{J\cos(\theta)-M}{\sin(\theta)}\right)\\
    &\times\cos(\theta/2)^{J+M}\sin(\theta/2)^{J-M}.
\end{align}
One can therefore calculate the locations of the discontinuities in the Fisher information by finding where
\begin{equation}
    h_M(\theta-\beta)=0,
\end{equation}
from which one finds that they occur at
\begin{equation}
    \theta-\beta=\pm\arccos(M/J).\label{eq:disc_point}
\end{equation}
By plugging these points back into our expression of $\hat\rho_\theta$, one can calculate using Eq.~(\ref{eq:disc_main}) that the value of the discontinuity is 
\begin{align}
    \Delta \mathcal F_C=&4|\mel{J,M}{e^{i\beta'\hat J_y}\hat J_ye^{-i\theta' \hat J_y}}{J,J-1}|^2\\
    =&8J\left(\frac{J+M}{2J}\right)^{J+M}\left(\frac{J-M}{2J}\right)^{J-M}\binom{2J}{J-M}\\
    \approx&8J\sqrt{\frac{J}{\pi(J+M)(J-M)}},\label{eq:disc_size}
\end{align}
where $\theta'$ and $\beta'$ in the above expression are chosen such that Eq.~(\ref{eq:disc_point}) is satisfied. The approximation in the final line is the normal approximation to the binomial distribution for large $J$. 

From Eq.~(\ref{eq:disc_size}) we notice that some discontinuities will become negligible in the limit as $J\to\infty$. Measurements where $\theta-\beta\approx\pi/2$ (which correspond to measuring along an axis orthogonal to the initial state) have discontinuities occurring from some outcome $M\notin \mathcal S_\theta$ for $M\ll J$. Therefore such measurements have discontinuities which scale as
\begin{equation}
    \Delta \mathcal F_C\sim\sqrt J.\label{eq:disc_size_small_m}
\end{equation}
In general, if $|M|=\alpha J$ for some rational constant $\alpha$, then we find that
\begin{equation}
    \Delta \mathcal F_C=\frac{8\sqrt J}{\sqrt{\pi(1-\alpha^2)}}\sim\sqrt J.\label{eq:disc_cfi_scaling}
\end{equation}
When not operating at points of discontinuity, the total CFI scales with $\sim J$. Therefore the ratio of Eq.~(\ref{eq:disc_cfi_scaling}) to the total Fisher information of nearby states scales as $\sim\sqrt J/J=1/\sqrt J$. This means that discontinuities formed from $M\notin\mathcal S_\theta$ with $|M|=\alpha J$ tend to be negligible as $J\to\infty$.

However, when measuring along an axis approximately aligned with the initial state $\theta-\beta\approx0$ the nearby discontinuities are caused by orthogonality to $M\approx J-n$ where $n\ll J$ is some constant. With this scaling plugged into Eq.~(\ref{eq:disc_size}) we find that
\begin{equation}
    \Delta\mathcal F_C\approx4J\sqrt{\frac{2}{\pi n}}\sim J,
\end{equation}
meaning these discontinuities remain on the order of the total Fisher information for large $J$. This means that if one attempts to measure along an axis which is aligned with the initial state, such that $\theta-\beta\approx\arccos((J-n)/J)\approx\sqrt{2n/J}$, then discontinuities remain on the order of the Fisher information of non-discontinuous states. Thus, these discontinuities do not become negligible in the large $J$ limit.



\medskip

\section{Minimizing Ensemble Decomposition}
\label{sec:min_ens_dec}
In this section, we will provide some more details on the proof provided in Sec.~\ref{sec:general_noise}, giving a proof that Jensen's inequality is saturated if and only if the score functions of all distributions in the ensemble are equal on their shared support. The condition derived in that section for Jensen's inequality to be saturated is 
\begin{align}
    &\frac{\mathrm d^{k}}{\mathrm d^kw_i}\mathcal F[\hat\rho_\theta-w_i\hat\sigma_\theta^{(j)},\mathcal M]\\
    &=\frac{\mathrm d^{k}}{\mathrm d^kw_i}\sum_\lambda \frac{(\partial_\theta p_\theta(\lambda)-w_i \partial_\theta g_\theta(\lambda))^2}{p_\theta(\lambda)-w_i g_\theta(\lambda)}=0,
\end{align}
for $k\geq2$. This simplifies to
\begin{align}
    \frac{\mathrm d^{k}}{\mathrm d^kw_i}&\mathcal F[\hat\rho_\theta-w_i\hat\sigma_\theta^{(j)},\mathcal M]\\
    \nonumber=&k!\sum_\lambda\left(\frac{g_\theta(\lambda)}{p_\theta(\lambda)-w_ig_\theta(\lambda)}\right)^k\\
    &\times\frac{\left(p_\theta(\lambda)\partial_\theta g_\theta(\lambda)-g_\theta(\lambda)\partial_\theta p_\theta(\lambda)\right)^2}{g_\theta(\lambda)^2(p_\theta(\lambda)-w_ig_\theta(\lambda))}.\label{eq:solved_derivative}
\end{align}
We notice that when $k=2$, the above expression contains only positive terms in the summand, and thus can only be equal to zero if for all $\lambda$
\begin{equation}
    p_\theta(\lambda)\partial_\theta g_\theta(\lambda)-g_\theta(\lambda)\partial_\theta p_\theta(\lambda)=0. \label{eq:score_condition}
\end{equation}
Furthermore, if the above condition is satisfied, then Eq.~(\ref{eq:solved_derivative}) equals zero for all $k>2$. Therefore, satisfying Eq.~(\ref{eq:score_condition}) is both necessary and sufficient for saturating Jensen's inequality.

\section{Details on Approximate Loss of Fisher Information}
\label{sec:appx_loss_details}

In this section, we will provide a more detailed calculation of the approximate loss of Fisher information for states near fragile probe states. We begin by writing an approximation of the amount of CFI lost after a small amount of noise is applied over a time $\gamma\Delta t$ such that the probability distribution is $P_\theta(\lambda,\Delta t)=p_\theta(\lambda)+\gamma\Delta t\sigma_\theta(\lambda)$, where
\begin{align}
    \sum_{\lambda}\sigma_\theta(\lambda)=0.
\end{align}

For this distribution we write that the change in the CFI over a small time $\gamma\Delta t$ is
\begin{widetext}
\begin{align}
    \nonumber\mathcal F_C[P_\theta(\Lambda,0);\theta]&-\mathcal F_C[P_\theta(\Lambda,\Delta t);\theta]\\
    &=\sum_\lambda\frac{(\partial_\theta p_\theta(\lambda))^2}{p_\theta(\lambda)}-\frac{(\partial_\theta p_\theta(\lambda)+\gamma\Delta t\partial_\theta\sigma_\theta(\lambda))^2}{p_\theta(\lambda)+\gamma\Delta t\sigma_\theta(\lambda)}\\
    &=\sum_\lambda \frac{\frac{(\partial_\theta p_\theta(\lambda))^2}{p_\theta(\lambda)}(p_\theta(\lambda)+\gamma\Delta t\sigma_\theta(\lambda))-(\partial_\theta p_\theta(\lambda)+\gamma\Delta t\partial_\theta\sigma_\theta(\lambda))^2}{p_\theta(\lambda)+\gamma\Delta t\sigma_\theta(\lambda)}\\
    &\approx\sum_\lambda \frac{(\partial_\theta p_\theta(\lambda))^2+\gamma\Delta t(\sigma_\theta(\lambda)/p_\theta(\lambda))(\partial_\theta p_\theta(\lambda))^2-(\partial_\theta p_\theta(\lambda))^2-2\gamma\Delta t(\partial_\theta p_\theta(\lambda))(\partial_\theta \sigma_\theta(\lambda))}{p_\theta(\lambda)+\gamma\Delta t\sigma_\theta(\lambda)}\\
    &=\sum_\lambda \frac{\frac{(\partial_\theta p_\theta(\lambda))^2}{p_\theta(\lambda)}-2(\partial_\theta p_\theta(\lambda))(\partial_\theta \sigma_\theta(\lambda))/\sigma_\theta(\lambda)}{\frac1{\gamma\Delta t\sigma_\theta(\lambda)}p_\theta(\lambda)+1}.\label{eq:cfi_before_appx}
\end{align}
\end{widetext}
Next, we note that there are two important regimes under which we can consider Eq.~(\ref{eq:cfi_before_appx}). When $p_\theta(\lambda)\gg\gamma\Delta t$, the denominator is small and thus the summand is zero. However, when $p_\theta(\lambda)\sim\gamma\Delta t\ll1$ the summand is nonzero and thus we may approximate both cases by
\begin{align}
    &\sum_\lambda\frac{\frac{(\partial_\theta p_\theta(\lambda))^2}{p_\theta(\lambda)}-2(\partial_\theta p_\theta(\lambda))(\partial_\theta \sigma_\theta(\lambda))/\sigma_\theta(\lambda)}{\frac1{\gamma\Delta t\sigma_\theta(\lambda)}p_\theta(\lambda)+1}\nonumber\\
    &\hspace{1.5in}\approx\sum_\lambda\frac{\frac{(\partial_\theta p_\theta(\lambda))^2}{p_\theta(\lambda)}}{\frac1{\gamma\Delta t\sigma_\theta(\lambda)}p_\theta(\lambda)+1}\\
    &\hspace{1.5in}\approx\sum_\lambda\frac{\Delta\mathcal F_C^{(\lambda)}[p_\theta(\Lambda);\theta]}{\frac1{\gamma\Delta t\sigma_\theta(\lambda)}p_\theta(\lambda)+1},\label{eq:robustness}
\end{align}
where in the last line we have approximated the numerator as the size of the discontinuity $\Delta \mathcal F_C^{(\lambda)}$ in the Fisher information associated with $\lambda\in\mathcal N_\theta$ since we assumed $p_\theta(\lambda)$ is vanishingly small. It's important to note that our assumption that $p_\theta(\lambda)\ll1$ cannot simultaneously be true for all $\lambda$ in the sums above. However, because for $p_\theta(\lambda)\gg\gamma\Delta t$ the summand approaches 0, we may keep the sum over all $\lambda$ and the expression remains approximately correct up to these terms of order $\gamma\Delta t$.

\section{Explicit Construction of Pathological Jump Operators}
\label{sec:explicit_construction}
In this section we will explicitly describe the construction of the operators $\hat L_{M}$ used in Fig.~\ref{fig:removed_discontinuities} and described in Sec.~\ref{sec:markovian_noise}. These operators satisfy Eq.~(\ref{eq:disc_special_case}) for $\ket\psi=\ket{N/2,N/2-1}$ with $N=32$, and $\ket{J,M}$ an eigenstate of $\hat J_z$. The unitary applied to the POVM is $\hat U_\beta=\exp(-i\beta \hat J_y)$. We define $\ket{M}_\beta\equiv\hat U_\beta\ket{J,M}$ such that $\bra{M}_\beta\ket\psi=0$ and construct a family of orthonormal states $\{\ket{\psi_i}\}$ with $\ket{\psi_0}=\ket\psi$. Using these we define an orthonormal operator basis
\begin{equation}
    \hat B_{M,i}=\ket{M}_\beta\bra{\psi_i}.
\end{equation}
Using this we can define jump operators which behave similarly to some arbitrary operator $\hat A$, but which satisfies Eq.~(\ref{eq:disc_special_case}). For example, in Sec.~\ref{sec:markovian_noise} we construct the jump operator $\hat L_{M}$ as
\begin{equation}
    \hat L_{M}=\sum_{M',i}\ket{M'}_\beta\bra{\psi_i}\bra{M'}_\beta \hat J_z\ket{\psi_i}(1-\delta_{M,M'}\delta_{i,0}).
\end{equation}
Without the Kronecker-deltas this expression would simply be $\hat J_z$, but the delta functions ensure that it does not couple $\ket\psi$ to $\ket{M}_\beta$, thus satisfying Eq.~(\ref{eq:disc_special_case}).

\end{document}